\documentclass[10pt]{iopart}

\usepackage{iopams}
\usepackage[T1]{fontenc}
\usepackage[utf8]{inputenc}
\usepackage{graphicx}
\usepackage{tabularx}
\usepackage{color, soul}
\usepackage{textcomp}
\usepackage{enumerate}
\usepackage{todonotes}
\usepackage[draft]{hyperref}
\usepackage{comment}
\usepackage{subcaption}
\usepackage{url}
\usepackage[normalem]{ulem}

\renewcommand{\vec}[1]{\mathbf{#1}}

\providecommand{\RV}[1]{#1}

\begin{document}

\title[Comparing simulations and experiments of positive streamers in air]{Comparing simulations and experiments of positive streamers in air: steps toward model validation}

\author{Xiaoran Li$^{1,2}$, Siebe Dijcks$^{3}$, Sander Nijdam$^3$, Anbang Sun$^2$, Ute Ebert$^{1,3}$, Jannis Teunissen$^1$}

\address{$^1$Centrum Wiskunde \& Informatica, Amsterdam, The Netherlands\\
	$^2$State Key Laboratory of Electrical Insulation and Power Equipment, School of Electrical Engineering,
	Xi'an Jiaotong University, Xi'an, 710049, China\\
  $^3$ Eindhoven University of Technology, Eindhoven, The Netherlands}
\ead{jannis.teunissen@cwi.nl}
\vspace{10pt}
\begin{indented}
\item[]\today
\end{indented}

\begin{abstract}
  We compare simulations and experiments of \RV{single} positive streamer discharges in air
  at 100\,mbar, aiming towards model validation. Experimentally, streamers are
  generated in a plate-plate geometry with a protruding needle. We are able to
  capture the complete time evolution of reproducible single-filament streamers
  with a ns gate-time camera. A 2D axisymmetric drift-diffusion-reaction fluid
  model is used to simulate streamers under conditions closely matching those of
  the experiments. Streamer velocities, radii and light emission profiles are
  compared between model and experiment. Good qualitative agreement is observed
  between the experimental and simulated optical emission profiles, and for the
  streamer velocity and radius \RV{during the entire evolution}. Quantitatively, the simulated streamer velocity
  is about 20\% to 30\% lower at the same streamer length, and the simulated
  radius is about 1\,mm (20\% to 30\%) smaller. The effect of various parameters
  on the agreement between model and experiment is studied, such as the used
  transport data, the background ionization level, the photoionization rate, the
  gas temperature, the voltage rise time and the voltage boundary conditions. An
  increase in gas temperature due to the 50\,Hz experimental repetition frequency
  could probably account for some of the observed discrepancies.
\end{abstract}

\ioptwocol

\section{Introduction}
\label{sec:introduction}

Streamer discharges are a common initial stage of electrical discharges
consisting of weakly ionized channels. 
The elongated shape of these channels greatly enhances the electric field at
their tips, which causes rapid growth of the channels there due to electron-impact
ionization. Positive streamers require a source of free electrons ahead of them. In air, photoionization is often the dominant source of such free electrons. In nature, streamers occur for example as sprites or as the precursors to lightning leaders~\cite{ebert2010}. They are also used in diverse technological applications~\cite{laroussi2014, bardos2010, popov2016,bruggeman2016}. A key property for most applications is the non-equilibrium nature of streamers. Due to their strong field enhancement, electrons can temporarily obtain energies of up to tens of eV while the background gas remains cold.

Streamer discharges have been extensively studied, both experimentally and
through modeling, see e.g.~\RV{the recent reviews~\cite{nijdam2020, babaeva2021}}.
  Numerical simulations are increasingly used to help explain experimental
  results and to study the physics of streamer discharges. Simulations provide the full temporal and spatial evolution of fields and plasma species, which are experimentally challenging to obtain. However, high computational costs are often a limiting factor. Simulations are therefore usually performed with plasma fluid models, which are less costly than more microscopic particle-based methods, see e.g.~\cite{nijdam2020}. \RV{For the same reason, the use of Cartesian or axisymmetric 2D models has been more common than that of 3D models. 
  While 3D simulations are now increasingly used to investigate problems such as streamer branching~\cite{marskar2020, teunissen2016, plewa2018a}.}


An important and still partially open question is how well commonly used
  streamer discharge models approximate physical reality. If simulations are not
  just used for qualitative understanding but also for quantitative predictions,
  the so-called verification and validation~\cite{roache1998} (V\&V) of
  simulation codes is required. Here \emph{verification} means ensuring the model
  equations are correctly solved, and \emph{validation} means ensuring the model is
  consistent with experimental results. Recently, steps towards the verification
  of six simulation codes were taken in~\cite{bagheri2018}.

In this paper, we take the steps towards model validation for streamer
  discharges, extending past validation work~\cite{pancheshnyi2005a, ono2020} that is discussed in more
  detail below. A summary of the approach taken in this paper is given below:
\begin{itemize}
  \item We experimentally generate stable and reproducible single positive
  streamers in air in a plate-plate geometry with a protruding needle. 
  \item With a camera with ns gate time, the time evolution of the streamers was captured in great detail, as well as the shape of the emission profiles.
  \item A 2D axisymmetric fluid model was used to simulate streamers under conditions closely matching those of the experiments, e.g. using the same applied voltage waveform, gas, and electrode geometry.
  \item The model includes light emission, and this light emission is processed to be directly comparable with the experimental observations.
  \item We perform quantitative comparisons of streamer velocities, radii and light emission profiles between model and experiment.
  \item The effect of various parameters on the agreement between model and experiment is studied, such as numerical convergence, transport data sources, background ionization levels, photoionization rates, gas temperatures, voltage rise times and voltage boundary conditions.
\end{itemize}


For the simulations, we use a drift-diffusion-reaction
type fluid model with the local field approximation, as described in section~\ref{sec:methods}. This model is commonly used to simulate streamer discharges~\cite{teunissen2017, bagheri2020a, francisco2021},
and the aim of the present paper is to take steps towards its validation. To understand how reliable simulations are, we first study the deviation between experimental and simulation results in section~\ref{sec:comparison}. Then we perform parameter studies to investigate possible sources of the observed differences in section~\ref{sec:discussion}.

\subsection{Past work}


Below, we first briefly present examples of past work in which streamer simulations and experiments were compared.

Pancheshnyi et al.~\cite{pancheshnyi2005a} experimentally investigated cathode-directed streamer discharges in synthetic
air in a pressure range of 300 to 760\,Torr and compared with axisymmetric fluid simulations. Deviations of up to 35\% were observed in the anode current and in the streamer velocity.
The companion papers of Briels~\cite{briels2008a} and Luque~\cite{luque2008} presented measurements and simulations of short positive and negative streamers in air at standard
temperature and pressure.
Komuro et al.~\cite{komuro2012} compared the simulated and experimental light emission for discharges in a pin-plate electrode geometry using streak images. Good agreement was achieved for the propagation of the primary streamer front, and secondary streamers were observed in both the experiments and simulations. In a related publication~\cite{Komuro_2013} the effect of the pulse rise time was investigated, and qualitative agreement was found for the streamer development in experiments and simulations. \RV{In~\cite{komuro2018}, they extended the comparison to the distribution of electron densities, and qualitative agreement was achieved.}
Eichwald et al.~\cite{eichwald2008} compared simulations and experiments of primary and secondary streamers in a point-plane positive corona discharge, focusing on the production of oxygen and nitrogen radicals. The experimental and simulated production of these radicals were found to be in qualitative agreement.
Nijdam et al.~\cite{nijdam2016} investigated the role of free electrons in the guiding of positive streamers in nitrogen--oxygen mixtures through a combination of experiments and 3D simulations, with the latter supporting the experimental observations.
\RV{Marode et al.~\cite{marode2016} studied diffuse discharges with a 2D fluid model and experiments. Similar light emission structures were recognized.
  A related paper~\cite{brisset2019} investigated the electric field distribution in diffuse discharges at high over-voltages, using both fluid simulations and experiments. Experimentally, spectral line ratios were used to determine the electric field.
Similar maximal electric field strengths were found, but several qualitative discrepancies were observed in the obtained field distributions.
In contrast, in a recent comparison of a fluid model and E-FISH measurements~\cite{zhu2021}, good agreement was found for the shape of the electric field profile but not for its peak amplitude.}
\RV{Furthermore, the light emission from discharges was compared between simulations and experiments for a glow-like discharge in~\cite{tholin2011} and for a conical discharge at high over-voltage in~\cite{pechereau2014a}. Good
agreement for the maximal discharge diameter and estimated velocity was obtained in~\cite{pechereau2014a}.}

\RV{Ono et al.~\cite{ono2020} have recently focused on} comparing experiments and simulations.
A single-filament streamer was generated from a pointed anode to a planar cathode in atmospheric-pressure air. Branching was suppressed by simultaneously generating four streamers from pointed electrodes placed around the central electrode.
The experimental light emission intensity, streamer diameter, cathode current were compared with 2D axisymmetric fluid simulations. Most of the main discharge features could be reproduced by the model but discrepancies were also observed. One reason for this could be that in the simulations a single hyperbolically shaped electrode was used to mimic the field created by the combined pointed electrodes. The streamer propagation velocity was then used to fit the tip radius of this hyperbolic electrode, whereas ideally it would be a parameter to validate.

Plasma jets are related to streamer discharges.
Yousfi et al.~\cite{yousfi2012} investigated the ionization wave dynamics of a low-temperature plasma jet with 1.5D fluid simulations and experiments. Similar ionization wave velocities were found both experimentally and numerically.
Hofmans et al.~\cite{hofmans2020} compared experimental measurements and 2D axisymmetric fluid simulations of a kHz atmospheric pressure He plasma jet. Excellent agreement was obtained for the
gas mixture distribution, the discharge length and velocity and the electric
field in the discharge front. Based on this, Viegas et al.~\cite{Viegas_2020}
studied the interaction of a plasma jet with grounded and floating metallic
targets both experimentally and computationally.

Most of the studies mentioned above found qualitative agreement between
simulations and experiments. For a quantitative comparison one challenge is that
branching streamers are expensive to simulate, and that due to their stochastic
nature a statistical comparison is required. Conversely, generating stable
single streamers is difficult experimentally, as illustrated by the work
of~\cite{ono2020}. One of the novel aspects of this paper is that we are able to
generate such stable and reproducible streamers in a relatively simple geometry,
also suitable for simulations.

Finally, we also list several studies in which different streamer discharge models were compared. Li et al.~\cite{li2012a} have compared 3D particle, fluid and hybrid simulations for negative streamers in air without photoionization in overvolted gaps. We should point out that the classical fluid model, which is also used in the present paper, was not implemented correctly in this comparison.
Markosyan et al.~\cite{markosyan2015} evaluated the performance of three plasma fluid models: a first and second order drift-diffusion-reaction model based on respectively the local field approximation and the local energy approximation, and a high order fluid model by Dujko et al.~\cite{dujko2013}.
They compared these three models to a particle-in-cell/Monte Carlo (PIC/MC) code in 1D.
Bagheri et al.~\cite{bagheri2018} compared six simulation codes for 2D axisymmetric positive streamer discharges from six different research groups. Four of these codes were self-implemented and two made use of COMSOL. All groups used the same fluid model with the same transport coefficients. With sufficiently fine grids and small time steps, good agreement was observed between several codes. The code used in this paper is among them.

\section{Experimental \& Simulation Methods}
\label{sec:methods}

\subsection{Experimental method}
\label{sec:exp-setup}

Since streamer discharges are a reaction of a gaseous medium to strong electric fields, having good control over both the field and the gas is essential.
We use a \RV{quasi-cylindrical} vessel \RV{(as shown in~\ref{sec:vessel})} with a diameter of 324\,mm and a height of 380\,mm for which the discharge operating pressure range is 1--1000\,mbar.

\begin{figure}
	\centering
	\includegraphics[width=1.0\linewidth]{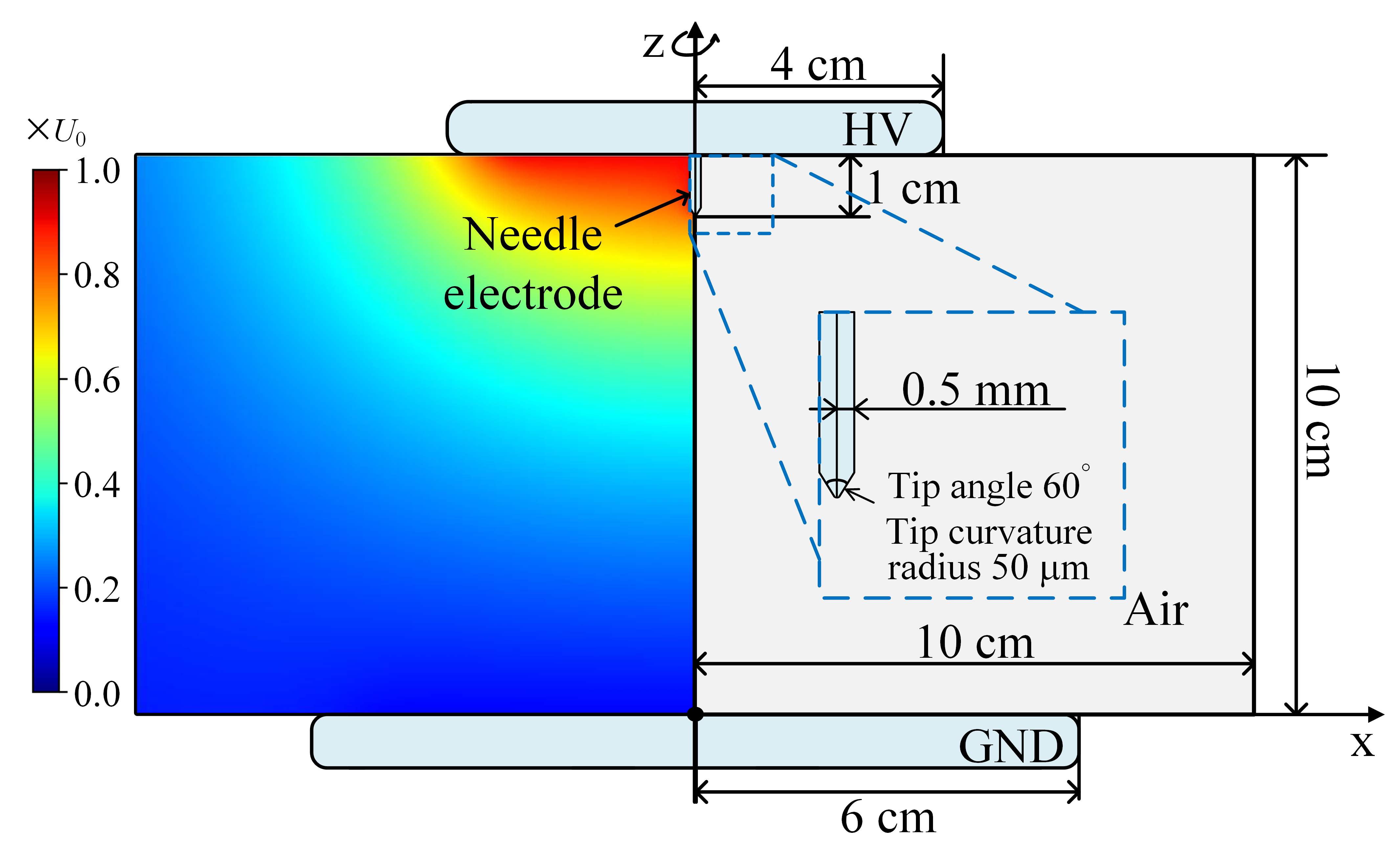}
	\caption{\RV{The electrode geometry in the experiments and simulations, consisting of parallel plates and a needle electrode from which discharges start. Right: the computational domain, for which $0~\leq~r,z~\leq 10$\,cm. The needle electrode is inside the computational domain and the plate electrodes are on its upper and lower boundary.  We use a coordinate system in which the electrode tip is at \mbox{$z=90 \, \textrm{mm}$} and the grounded plate electrode is at \mbox{$z=0 \, \textrm{mm}$}. Left: the electric potential in the absence of space charge.}
	}
	\label{fig:simulation-domain}
\end{figure}

The vessel is grounded and the electrode geometry inside it is illustrated in figure \ref{fig:simulation-domain}. An elevated grounded plate with a 6\,cm radius is positioned 10\,cm from the HV (high voltage) electrode, which has a 4\,cm radius. 
\RV{A 1\,cm long needle electrode with a 0.5\,mm radius is connected to the HV electrode.
This cylindrical
electrode ends in a cone with a 60$^{\circ}$ top angle that transitions into
a spherical tip with a radius of curvature of 50\,$\mu$m.}
The plate-plate geometry with a protruding needle results in a field that is approximately homogeneous in the gap, which suppresses streamer branching. The cylindrical symmetry of the vessel is broken at a distance of about 15\,cm from its center due to windows for optical access.

\begin{figure}
	\centering
	\includegraphics[width=0.9\linewidth]{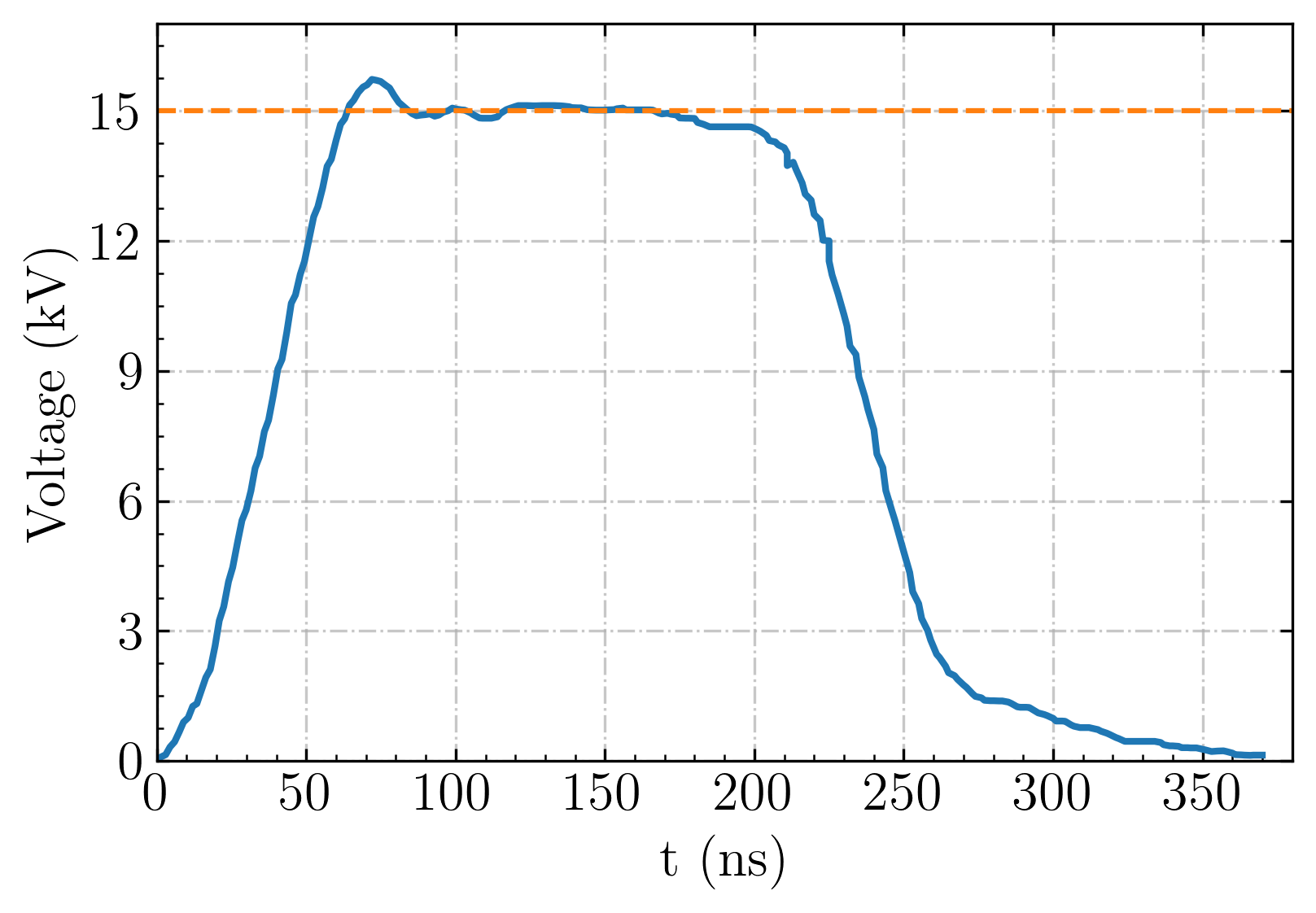}
	\caption{The voltage waveform as measured at the HV electrode. This waveform is also used in the simulations.}
	\label{fig:voltage_time}
\end{figure}

A strong field is generated at the protruding needle by applying a fast HV pulse. The high voltage is generated by a DC source (Spellman Bertan 205B), which charges a discharging capacitor (40\,kV | 2000\,pF), which in turn is discharged by a HV switch (Behlke HTS 651-15-Sic-GSM) coupling the charged capacitor to the HV electrode for 200\,ns at 50\,Hz with 350\,ps jitter on the start time. The voltage waveform at the HV electrode is shown in figure \ref{fig:voltage_time}.

Imaging is performed using an UV optimized ICCD (Lavision PicoStar HR\,+\,UV\,105\,mm lens) system. The CCD is synchronized with the discharge repetition rate, such that one discharge is imaged per exposure. The intensifier is then directly gated, where a gate of 900\,ps is sequentially delayed through the voltage waveform, creating a phase-resolved sequence of images depicting the propagation of the streamer. Each image has an effective resolution of about 0.2\,mm per pixel for the 10\,cm discharge gap, see~\ref{sec:appendix-a}. 
Most of the image intensity comes from the decay of excited nitrogen molecules in the plasma, with the second positive system contributing most, and smaller contributions from the first positive and negative systems.

\RV{With a 50\,Hz repetition rate remnants from previous discharges reduce the stochasticity in streamer inception~\cite{nijdam2011}. This greatly improves the stability of the discharges and thus the quality of the measurements. Besides inception, the propagation of consecutive discharges is essentially independent of that of previous ones at 50\,Hz~\cite{nijdam2011, nijdam2014}. Slight changes in frequency hardly affect streamer behavior, but changing the frequency by an order of magnitude leads to visually observable differences.}

All experiments were performed with the vessel at room temperature,
\RV{and a pressure controller kept the pressure inside the vessel at 0.1\,bar, with about 1\% uncertainty.}
The vessel was continuously flushed with 2\,SLM synthetic air while performing the experiments, giving a residence time of a couple of minutes.

\subsection{Simulation model}
\label{sec:sim-model}

We use a drift-diffusion-reaction type fluid model with the local field
approximation to simulate positive streamers in artificial air, composed of 80\% nitrogen and
20\% oxygen at 300\,K and 0.1\,bar.
Two-dimensional axisymmetric simulations are performed with \texttt{Afivo-streamer}~\cite{teunissen2017}, an open source code for the
plasma fluid simulation of streamer discharges.
The code is based on the Afivo framework~\cite{teunissen2018}, and it includes adaptive mesh refinement (AMR), geometric multigrid methods for Poisson’s equation and OpenMP parallelism. For a recent comparison of six streamer simulation codes, including \texttt{Afivo-streamer},
see~\cite{bagheri2018}.

The temporal evolution of the electron
density (\RV{$n_e$}) is given by
\RV{
\begin{equation}
	\partial_t n_e = \nabla \cdot (n_e \mu_e \vec{E} + D_e \nabla n_e) + S_i + S_\mathrm{ph} - S_\mathrm{attach},
\end{equation}
}
where $\mu_e$ is the electron mobility, $D_e$ the
electron diffusion coefficient and $\vec{E}$ the electric field. Furthermore, $S_i$, $S_\mathrm{attach}$ and $S_\mathrm{ph}$ are the respective source terms for impact ionization, attachment and non-local photoionization.
Photoionization is computed according to Zheleznyak's model~\cite{zheleznyak1982}, with the source of ionizing photons given by
\begin{equation}
  \label{eq:photoi-source}
  I_\mathrm{ph} = \frac{p_q}{p + p_q} \xi S_i,
\end{equation}
where $p$ is the gas pressure, \mbox{$p_q=40 \, \textrm{mbar}$} the quenching pressure, and $\xi$ a proportionality factor, which is here set to $0.075$~\cite{teunissen2016, zheleznyak1982}. The effect of $\xi$ is investigated in section~\ref{sec:photonionization}. Note that the factor \mbox{$p_q/(p+p_q)$} is about $7.4$ times larger at $0.1 \, \textrm{bar}$ than at $1 \, \textrm{bar}$, so that there is significantly more photoionization at $0.1 \, \textrm{bar}$.
The absorption of the ionizing photons is computed using the Helmholtz approximation with Bourdon's three-term expansion for the absorption function, as described in~ \RV{\cite{bourdon2007, bagheri2019}}. Ions are assumed to be immobile. The electric field $\vec{E}$ is calculated by solving Poisson's equation:
\begin{eqnarray}
	\nabla \cdot \left(\varepsilon_0 \nabla \varphi \right)=-\rho,\\
	\vec{E}=-\nabla \varphi,
\end{eqnarray}
where $\varepsilon_0$ is the vacuum permittivity and $\rho$ is the space charge density.

\subsubsection{Reactions and light emission}
The reactions considered in this paper are listed in table \ref{tbl:reaction_table}, including electron impact ionization ($k_1$ - $k_3$),
electron attachment ($k_4$, $k_5$) and reactions related to light emission ($k_6$ - $k_9$).
According to table \ref{tbl:reaction_table}, the impact ionization $S_i$ and the electron attachment source terms $S_\mathrm{attach}$ are calculated as,
\begin{eqnarray}
	S_i = n_\textrm{e} [\textrm{N}_2] k_1 + n_\textrm{e} [\textrm{N}_2] k_2 + n_\textrm{e} [\textrm{O}_2] k_3,\\
	S_\mathrm{attach} = n_\textrm{e} [\textrm{O}_2]^2 k_4 + n_\textrm{e} [\textrm{O}_2] k_5.
\end{eqnarray}
where $[\textrm{N}_2]$ indicates the number density of N$_2$, the same for $[\textrm{O}_2]$, and $k_j$ , $j = 1, 2, ..., 5$ are the respective reaction rates.

\begin{table*}
	\centering
	\caption{Reactions included in the model, with reaction rates and
          references. The availability of the transport and reaction data used in this paper is described at the end of section \ref{sec:summary}.}
	\begin{tabular}{cccc}
		\hline
		Reaction No. & Reaction & Reaction rate coefficient  & Reference \\ \hline
		1 & $\textrm{e} + \textrm{N}_2 \stackrel{k_1}{\longrightarrow} \textrm{e} + \textrm{e} + \textrm{N}_2^+$ (15.60 eV) & $k_1(E/N)$ &~\cite{phelps1985, hagelaar2005} \\
		2 & $\textrm{e} + \textrm{N}_2 \stackrel{k_2}{\longrightarrow} \textrm{e} + \textrm{e} + \textrm{N}_2^+$ (18.80 eV) & $k_2(E/N)$ &~\cite{phelps1985, hagelaar2005} \\
		3 & $\textrm{e} + \textrm{O}_2 \stackrel{k_3}{\longrightarrow} \textrm{e} + \textrm{e} + \textrm{O}_2^+$ & $k_3(E/N)$ &~\cite{phelps1985, hagelaar2005} \\
		4 & $\textrm{e} + \textrm{O}_2 + \textrm{O}_2 \stackrel{k_4}{\longrightarrow} \textrm{O}_2^- + \textrm{O}_2$ 	& $k_4(E/N)$ &~\cite{phelps1985, hagelaar2005}  \\
		5 & $\textrm{e} + \textrm{O}_2 \stackrel{k_5}{\longrightarrow} \textrm{O}^- + \textrm{O}$ 			& $k_5(E/N)$ &~\cite{phelps1985, hagelaar2005}  \\
		6 & $\textrm{e} + \textrm{N}_2 \stackrel{k_6}{\longrightarrow} \textrm{e} + \textrm{N}_2(\textrm{C}^3 \Pi_u)$ 	& $k_6(E/N)$ &~\cite{phelps1985, hagelaar2005} \\
		7 & $\textrm{N}_2(\textrm{C}^3 \Pi_u) + \textrm{N}_2 \stackrel{k_7}{\longrightarrow} \textrm{N}_2 + \textrm{N}_2$ & $k_7 = 0.13\times10^{-16}\,\textrm{m}^{3}\textrm{s}^{-1}$ &~\cite{pancheshnyi2005a} \\
		8 & $\textrm{N}_2(\textrm{C}^3 \Pi_u) + \textrm{O}_2 \stackrel{k_8}{\longrightarrow} \textrm{N}_2 + \textrm{O}_2$ & $k_8 = 3.0\times10^{-16}\,\textrm{m}^3\textrm{s}^{-1}$ &~\cite{pancheshnyi2005a} \\
		9 & $\textrm{N}_2(\textrm{C}^3 \Pi_u)\stackrel{k_9}{\longrightarrow} \textrm{N}_2(\textrm{B}^3 \Pi_g)$ 		& $k_9=1/(42\,\textrm{ns})$ &~\cite{pancheshnyi2005a} \\
		\hline
	\end{tabular}
	\label{tbl:reaction_table}
\end{table*}

To compare with the experimental observations, light emission of the second
positive system of nitrogen is modeled. The corresponding \mbox{
$\textrm{N}_2(\textrm{C}^3 \Pi_u \rightarrow \textrm{B}^3 \Pi_g)$} transition is the main source of emitted
light for nanosecond discharges in \mbox{$\textrm{N}_2-\textrm{O}_2$} mixtures around atmospheric
pressure~\cite{pancheshnyi2000a}.
In table 1, $k_6$ is the electronic excitation rate of the \mbox{$\textrm{N}_2(\textrm{C}^3 \Pi_u)$} level from the ground state; $k_7$ and $k_8$ are the quenching rate constants for N$_2$ and O$_2$, respectively; the radiative lifetime of \mbox{$\textrm{N}_2(\textrm{C}^3 \Pi_u)$} is \mbox{$1/k_9$ = 42\,ns}~\cite{pancheshnyi2005a}.

All transport and reaction coefficients ($k_1 - k_6$) depend on the reduced electric field $E/N$,
and they were computed using BOLSIG+~\cite{BOLSIG+} with Phelps' cross sections for \mbox{(N$_2$, O$_2$)}~\cite{phelps1985, Phelps-database}. In section~\ref{sec:transport_data} the effect of different cross sections and Boltzmann solvers is compared.

\subsubsection{Computational domain \& simulation conditions}
\label{sec:computational-domain}

The axisymmetric computational domain shown in figure~\ref{fig:simulation-domain} \RV{(the grey square)} was designed to closely resemble the experimental geometry.
\RV{The domain consists of the region $0~\leq~r,z~\leq~10$\,cm, which covers the gap bounded by the plate electrodes.
  As in the experiments, a 1\,cm long needle electrode is inserted at the HV electrode, with a 0.5\,mm radius.
  The electrode tip is a cone with a 60$^{\circ}$ top angle that ends in a spherical tip with a radius of curvature of about 50\,$\mu$m, just as in the experiments.
The potential at the contour of the needle electrode is fixed at the applied voltage, which was implemented by modifying the multigrid methods in \cite{teunissen2018} using a level-set function.}

In the radial direction, the domain extends up to 10 cm, which is less than the vessel radius (16.2\,cm).
\RV{The effect of the finite plate electrodes is simulated by using pre-computed Dirichlet boundary conditions on the upper and lower boundaries.
These boundary conditions were obtained by solving for the electric potential in the entire discharge vessel in the absence of a discharge, using a finite element model.
The resulting potential is shown in figure~\ref{fig:simulation-domain}.
Homogeneous Neumann boundary conditions are applied for the electric potential in the radial direction.
However, these boundary conditions may not closely match the experiments, but it is hard to use more correct ones. More information about how boundary conditions affect the results is given in section~\ref{sec:boundary_conditions}.
In the presence of a discharge the potential distribution at the upper and lower domain boundaries changes, but computational experiments showed that these changes were not significant. For simplicity, we therefore keep the potential profile at the top and bottom boundary fixed. These profiles are normalized and scaled with the actual applied voltage on the HV electrode ($U_0$), so that we can account for the voltage rise time.
In section~\ref{sec:finite-plate-electrode}, we study how the size of the plate electrodes affects streamer properties.}

\RV{For all plasma species densities, homogeneous Neumann boundary conditions are used on all the domain boundaries. At the needle electrode electron fluxes are absorbed but not emitted, and secondary electron emission was not taken into account since a positive voltage was applied.}

\RV{This is the first time we employ a needle electrode in \texttt{Afivo-streamer}. In previous computational studies, an elongated ionized seed was often used as a pseudo-electrode to start a streamer, see e.g.~\cite{bagheri2020a, xiaoran-psst-2020}. We compared the differences between starting a streamer with an electrode and with an initial ionized seed in~\ref{sec:electrode-vs-seed}.}


The conditions used for the discharge simulations are summarized in table \ref{tab:conditions}. 
In particular, the initial density of electrons and positive ions is set to $10^{11}$~m$^{-3}$. In the simulations the same applied voltage is used as in the experiments, as shown in figure \ref{fig:voltage_time}. The voltage increases from 10\% to 90\% of its full amplitude (15\,kV) in about 52\,ns, so that the voltage rise time from zero to full amplitude is about 65\,ns.

\begin{table}
  \centering
    \caption{A summary of simulation conditions. The sections in which the respective
    parameters are varied are indicated. The parameter $c_0$ is used for grid refinement, see section \ref{sec:model-convergence}.}
  \begin{tabular}{l p{3.5cm} c}
  	\hline
    Parameter & Value & Section\\
    \hline
    Gas composition & 80\% N$_{2}$, 20\% O$_{2}$ & -\\
    Gas pressure & $0.1$\,bar & -\\
    Gas temperature & 300\,K & \ref{sec:gas-temperature}\\
    Applied voltage & 15\,kV, 65 ns rise time, see Fig. \ref{fig:voltage_time}  & \ref{sec:voltage-effect}\\
    Initial ionization & $10^{11}$\,m$^{-3}$ electrons and positive ions (uniform) & \ref{sec:background-densities} \\
    Numerical grid & \mbox{$\Delta x_{\mathrm{min}}$ = 6.1\,$\mu$m} \mbox{($c_0 = 0.5$)} & \ref{sec:model-convergence}\\
    \hline
  \end{tabular}
  \label{tab:conditions}
\end{table}

\subsection{Processing of emitted light}
\label{sec:processing-of-light}

Experimentally, the streamer morphology is captured with an ICCD camera.
To quantitatively compare the simulated streamers with experiments, it is important to accurately model the light emission from the discharge, and to process it in the same way for both the experiments and simulations.

As already mentioned above, the \mbox{$\textrm{N}_2(\textrm{C}^3 \Pi_u \rightarrow \textrm{B}^3 \Pi_g)$} transition is responsible for most of the optical emission under our discharge conditions~\cite{pancheshnyi2000a}.
Therefore the number of photons emitted at any given time is approximately 
proportional to the \mbox{$\textrm{N}_2(\textrm{C}^3 \Pi_u)$} density, which is included in the discharge model, see table~\ref{tbl:reaction_table}. Experimentally, we get a good approximation of the instantaneous light emission by using a short camera gate time of 900\,ps. As shown in section~\ref{sec:comparison}, typical streamer velocities under the present conditions are on the order of $0.5$ to $1\,\textrm{mm/ns}$, which means that the streamers move less than a\,mm during the camera gate time.

To compare the light from axisymmetric simulations with experimental observations, we have to apply a forward Abel transform. For this purpose, the \mbox{$\textrm{N}_2(\textrm{C}^3 \Pi_u)$} density in the region \mbox{$0 \leq r \leq 15$\,mm} by \mbox{$0 \leq z \leq 90$\,mm} (from the grounded electrode to the needle electrode tip) is first stored on a uniform grid, with a resolution \mbox{$\delta r$ = 0.01\,mm} and \mbox{$\delta z$ = 0.05\,mm}. The Hansen--Law method~\cite{hansen1985} is used for the forward Abel transformation.
The experimental pictures are cropped to the same region,
so that the light from the simulations and experiments is described by profiles \mbox{$I(x, z)$}, where \mbox{$z \in [0, 90] \, \mathrm{mm}$} is the propagation direction and \mbox{$x \in [-15, 15] \, \mathrm{mm}$} is the direction perpendicular to it.

\begin{figure}
	\centering
	\includegraphics[width=1.0\linewidth]{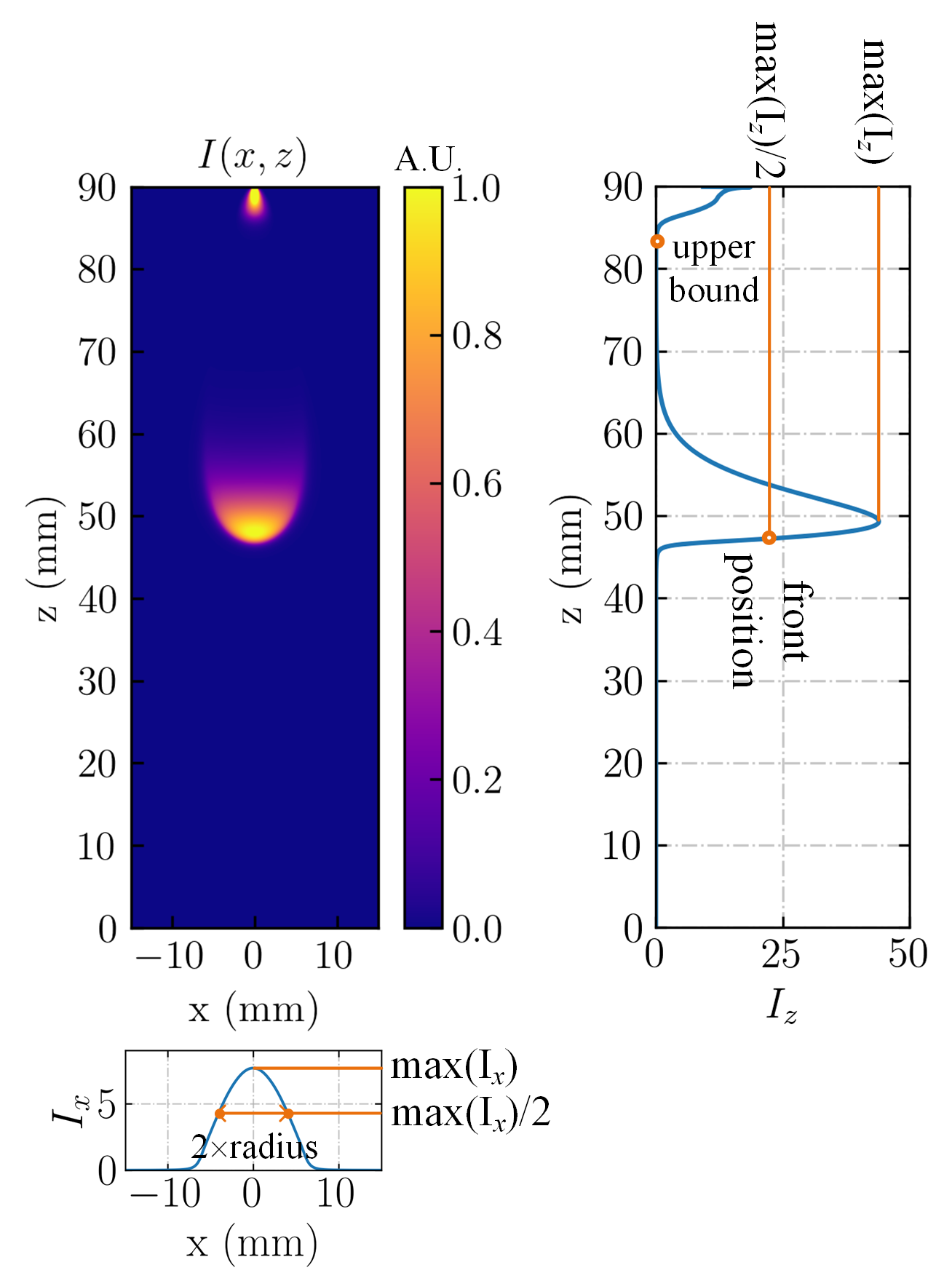}
	\caption{Illustration showing how the streamer front position and radius
          are determined from the light emission profile. The $z$ axis shows the 90\,mm between the tip of the (10\,mm long) needle electrode and the grounded plate electrode.}
	\label{fig:light-process}
\end{figure}

To directly compare streamer front positions, velocities and radii between experiments
and simulations, we determine these properties based on the emitted light. The procedure is illustrated in figure~\ref{fig:light-process}.
To obtain the streamer's front position, we first compute $$I_z(z) = \int_{-15 \, {\rm mm}}^{15 \, {\rm mm}} I(x, z) dx.$$
The front position is then determined as the minimum $z$
coordinate where $I_z(z)$ exceeds half of its maximum.
Streamer velocities are determined by taking the numerical time derivative of these $z$ coordinates for consecutive images.
For the radius we follow a similar approach, first computing
$$I_x(x) = \int_{0}^{z_\mathrm{ub}} I(x, z) dz.$$
The upper bound $z_\mathrm{ub}$ is used to exclude strong emission around the tip of the needle electrode. $I_x(x)$ therefore mostly consists of
light emitted close to the streamer head. The streamer optical radius is then defined as the FWHM (full width at half maximum) of $I_x(x)$. A similar definition has been used in earlier work, e.g.~\cite{briels2008a}.

\section{Comparison of emission profiles and streamer properties}
\label{sec:comparison}

\begin{figure*}
	\centering
	\includegraphics[width=0.8\linewidth]{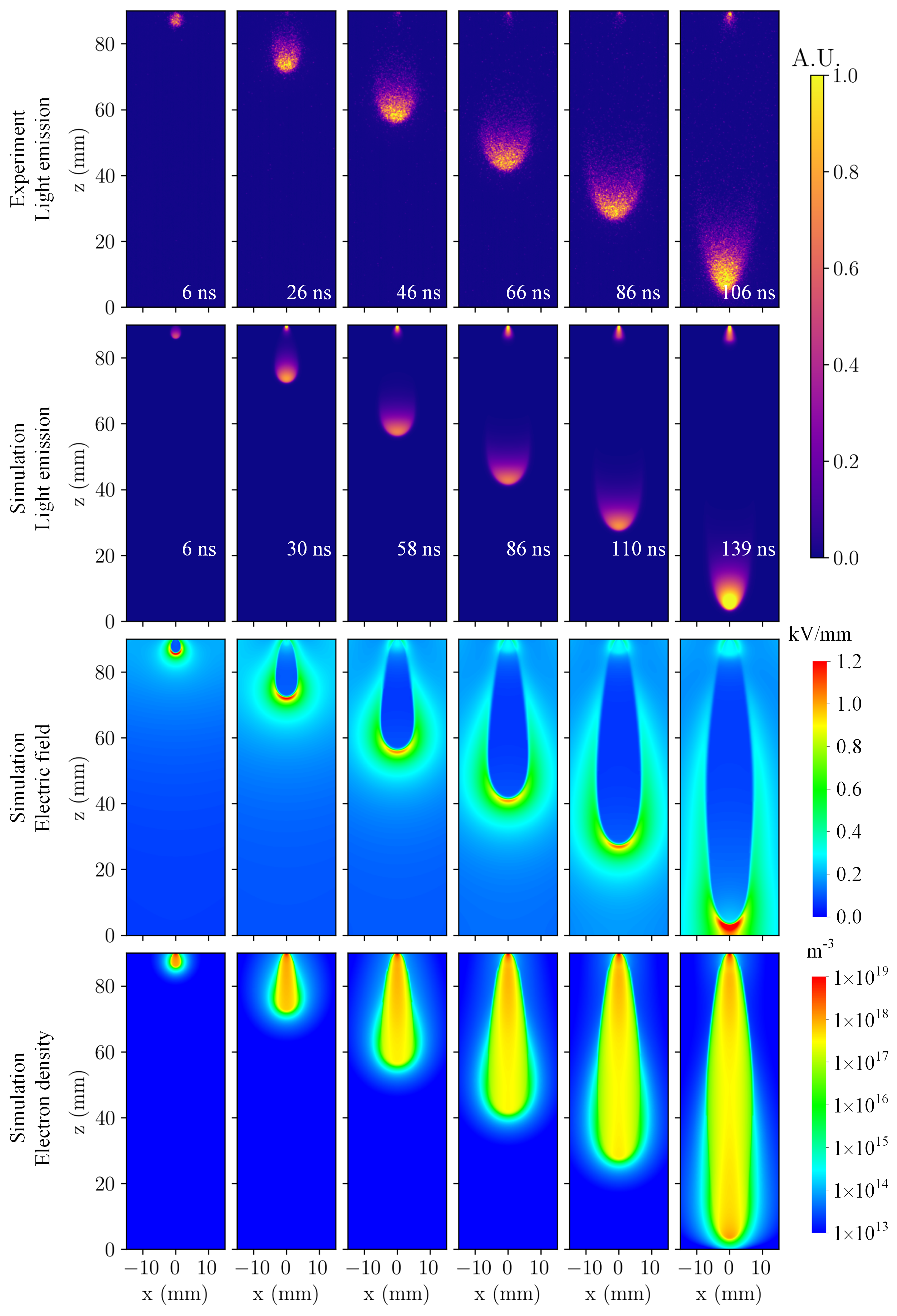} 
	\caption{From top to bottom: the light emission profile of experiments
          (camera gate time 900\,ps), the simulated instantaneous light emission
          profile, the simulated electric field and electron
          density.
          Each frame only shows part of the ICCD images/simulation domain.
          The $x$ axis shows $\pm$ 15\,mm around the center of the needle electrode.
          The $z$ axis shows the 90\,mm between the tip of the (10\,mm long) needle electrode and the grounded plate electrode.
          The experimental and simulated streamers in the same row have similar streamer lengths.
          The moment when the streamer length just exceeds 2\,mm is taken as 0\,ns.
          For light emission the data was normalized \emph{per row} to arbitrary units, so that frame-to-frame brightness variations are conserved.
          This was done by dividing by the value at their $0.999$\textsuperscript{th} quantile, and limiting the result to the range \mbox{[0, 1]}. This ensures that a few bright pixels do not affect the brightness of the streamer head.
          }
	\label{fig:big-figure}
\end{figure*}

Figure~\ref{fig:big-figure} shows the experimental and simulated light emission profile from 6\,ns to the last frame captured, together with the simulated electric field and electron density.
There is good qualitative agreement between the emission profiles, although the experimental streamer has a higher velocity and a larger radius.
In both the experimental and simulation figures the streamers' characteristic head shape is visible.
The front of the streamer heads is always the brightest, a bit like a crescent moon, which is followed by a darker tail due to the decay of the emitting \mbox{$\textrm{N}_2(\textrm{C}^3 \Pi_u)$} molecules.
The streamers grow wider as they propagate down, but when they approach the grounded electrode they accelerate, their radius reduces and their heads become even brighter. At the same time, the electric field and the electron density at the streamer head also increase.

\begin{figure}
	\centering
	\includegraphics[width=0.9\linewidth]{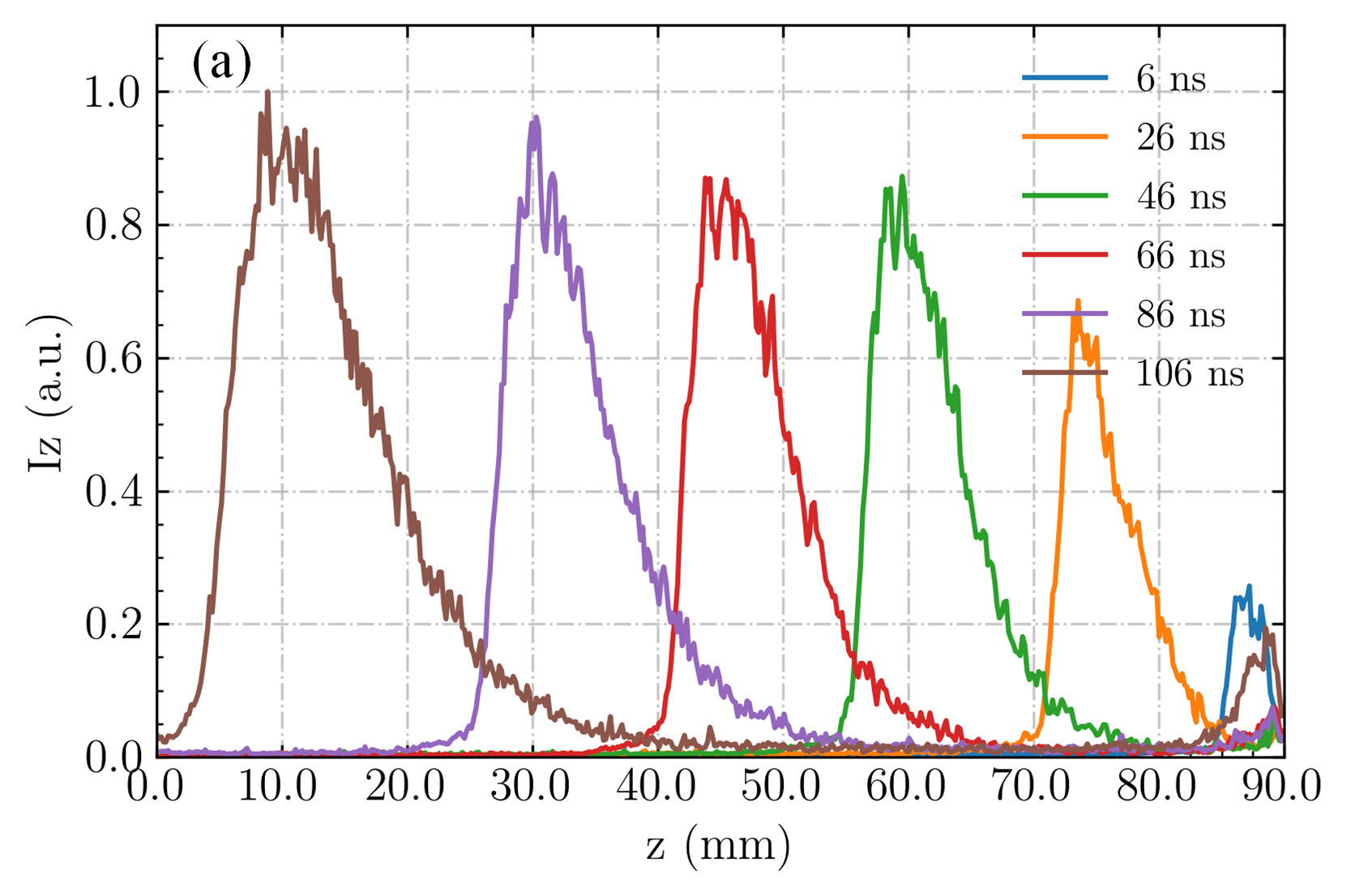} 
	\includegraphics[width=0.9\linewidth]{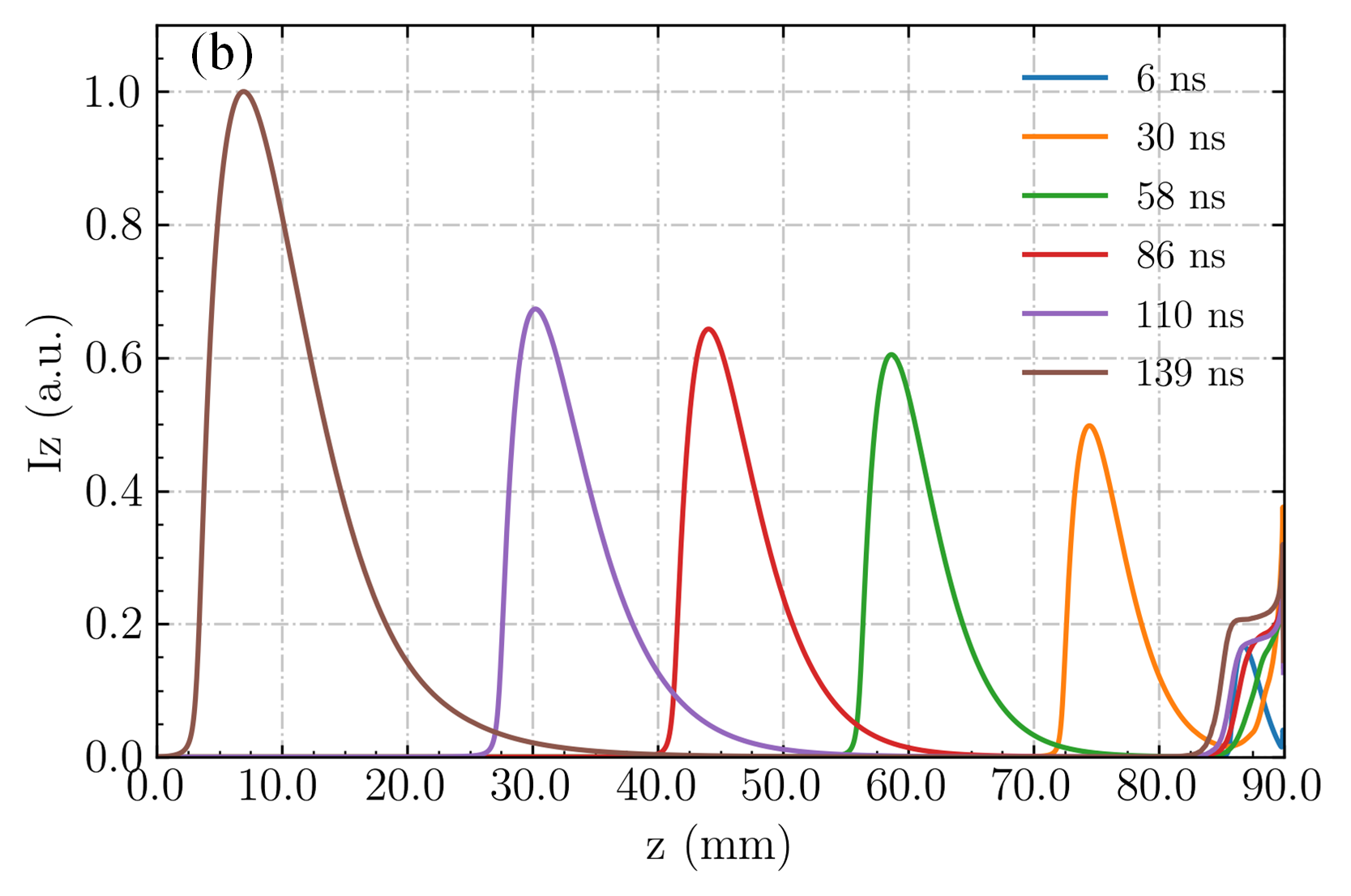}
	\caption{The integrated light emission profile $I_z$ for the experimental (a) and simulated (b) streamers in figure \ref{fig:big-figure}. In each sub-figure the data was normalized to a maximal amplitude of one.} 
	\label{fig:Iz}
\end{figure}

Figure \ref{fig:Iz} shows the integrated light emission profile $I_z$ for the experimental and simulated streamers in figure \ref{fig:big-figure}. When
  compared at the same length, most of the curves look similar. However, at the
  final time the amplitude of the simulated light emission is significantly
  larger. Another difference is that the tail of the emitted light is narrower
  in the simulations.

\begin{figure}[htb]
  \centering
  \includegraphics[width=0.9\linewidth]{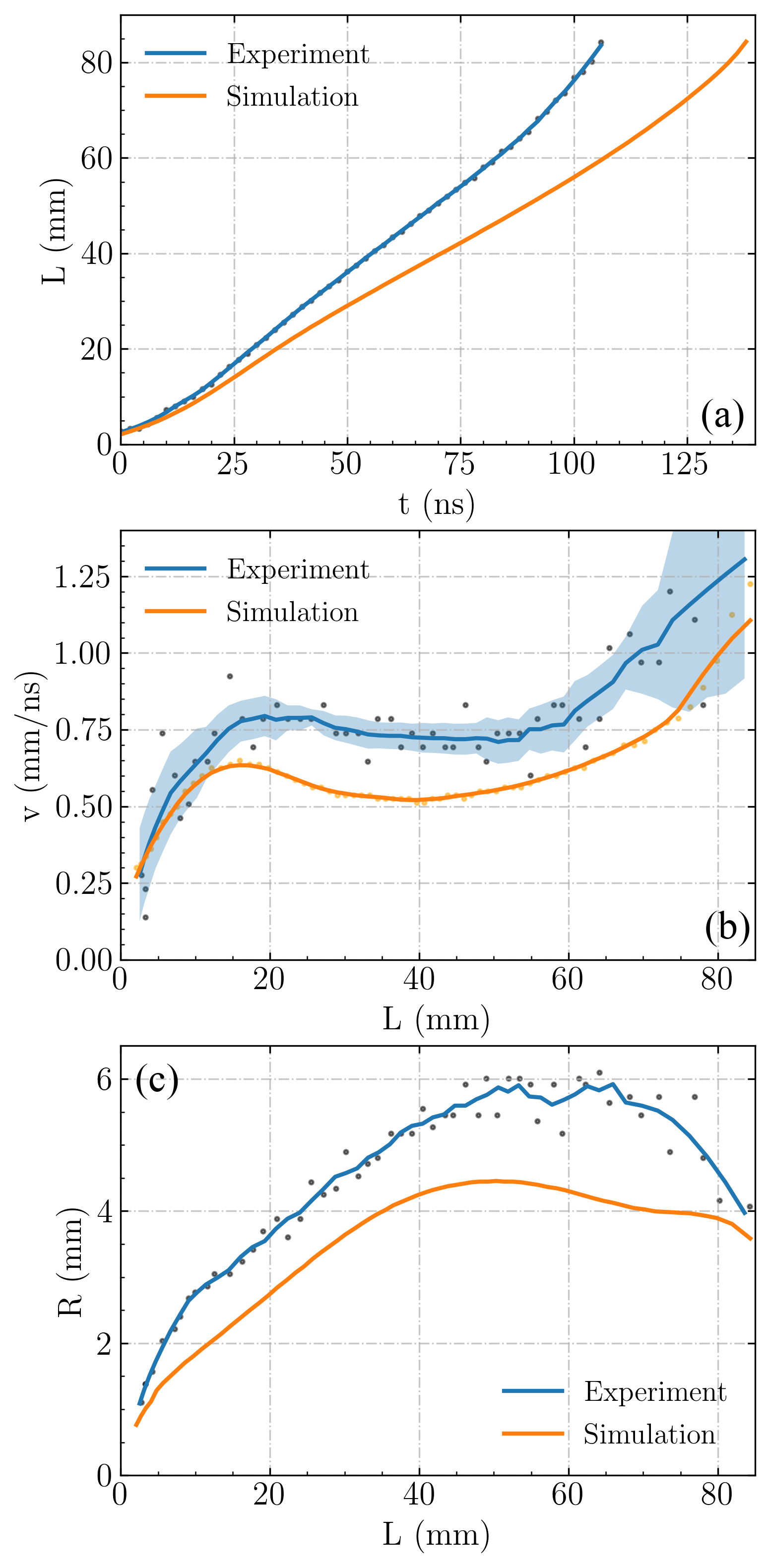}
  \caption{Comparison of streamer propagation parameters between
    experiments and simulations. (a) Streamer length versus time. (b) Streamer velocity versus
    streamer length. (c) Streamer radius versus streamer
    length. 
    The dots indicate unsmoothed data.
	The blue filled area shows the standard deviation between the unsmoothed and smoothed experimental velocity.}
  \label{fig:velocity-and-radius}
\end{figure}

Figure~\ref{fig:velocity-and-radius}(a) to (c) show the streamer length versus
time and the streamer velocity and radius versus the streamer length,
respectively. Experimentally, each measurement is obtained from a new discharge,
which leads to some fluctuations in the streamer properties. 
These fluctuations are smoothed by a second order Savitzky–Golay filter with a window size of
nine~\cite{schafer2011}.


Qualitatively, the agreement in the streamer velocity profile is quite good.
After inception, the streamers first accelerate and then they slowly decelerate. Afterwards, they obtain an approximately stable velocity, and finally they accelerate again when they approach the opposite electrode. All these phases are present in both the experimental and simulation data, although the times and streamer lengths at which they occur are somewhat different.
The maximal electric field at the streamer head follows a similar trend as the streamer velocity, as can be seen in figure~\ref{fig:big-figure}. The deceleration of the streamers in the middle of the gap is related to the size of the plate electrodes, as discussed in detail in section \ref{sec:finite-plate-electrode}.
There is also good qualitative agreement in the streamer radius between simulations and experiments. The radius initially increases until the streamers are about 50\,mm long, and then it decreases when the streamers approach the opposite electrode.

Quantitatively, figure~\ref{fig:velocity-and-radius} shows that the simulated streamer velocity is about 20\% to 30\% lower when compared at the same streamer length, and the simulated radius is about 1 - 1.2\,mm smaller (also 20\% to 30\%). 
These discrepancies could well be correlated, as earlier studies~\cite{briels2008a,luque2008,naidis2009} have found that the streamer velocity increases with the streamer radius. On the other hand, the observed streamer velocities do not increase with the radius for streamer lengths between 15\,mm and 40\,mm because the streamer's maximal field in this region decreases.

Going back to figure~\ref{fig:big-figure}, there is one detail in which the experimental and simulation results disagree: the emitted light near the electrode tip.
In the simulations, a bright spot is always visible, whereas in the experiments this only happens occasionally.
This could be related to the width of the streamer channel connected to the needle electrode, since a narrower connection means that a higher field and a higher electron density are required to carry the discharge current, leading to more light emission. These effects are visible in figure~\ref{fig:big-figure}, in which the electric field in this region is about $0.3$\,kV/mm and the electron density is about $1 \times 10^{19}$\,m$^{-3}$.

As discussed in section~\ref{sec:background-densities}, discharge inception is
sometimes not accurately described by a fluid model because the continuum
approximation breaks down when there are few particles. This could affect the
connection of the discharge to the electrode, and thereby also the light
emission around this area. Furthermore, the voltage rise time also affects the
brightness of this area, see section~\ref{sec:voltage-effect}.

\section{Investigating possible sources of discrepancy}
\label{sec:discussion}

The results in section~\ref{sec:comparison} showed good qualitative agreement
between the simulations and experiments. However, the simulated streamer velocity was 20\% to 30\% slower, and the streamer radius was about \mbox{1 - 1.2\,mm} (20\% - 30\%)
smaller, when compared at the same streamer length. In this section, we
investigate how several simulation and discharge parameters affect these
quantitative differences. Below we only mention the parameters that are changed,
all other parameters are set according to table~\ref{tab:conditions}.

\subsection{Numerical convergence}
\label{sec:model-convergence}

Model verification means checking whether the
model's equations are solved correctly and with sufficient numerical accuracy, which is an important step towards the development of validated models.
In an earlier study~\cite{bagheri2018} the \texttt{Afivo-streamer} code was
compared against five other codes for this purpose. It was found that with sufficiently fine grids and small time steps different codes could produce highly similar results, indicating numerical convergence. Below, we again test the numerical convergence of our model for the present discharge simulations.

For computational efficiency, \texttt{Afivo-streamer} uses adaptive mesh refinement (AMR).
The refinement criterion is based on 1/$\alpha(E)$, which is the average distance between ionization events for an electron~\cite{teunissen2017}:
\begin{itemize}
  \item refine if $\Delta x > c_0 / \alpha (E)$
  \item de-refine if $\Delta x < \mathrm{min}\{0.125 \, c_0 / \alpha (E),\, d_0\}$
\end{itemize}
where $\alpha (E)$ is the field-dependent ionization coefficient, $\Delta x$ is
the grid spacing, and $c_0$ and $d_0$ are constants.

\begin{figure}[htb]
	\centering
	\includegraphics[width=0.9\linewidth]{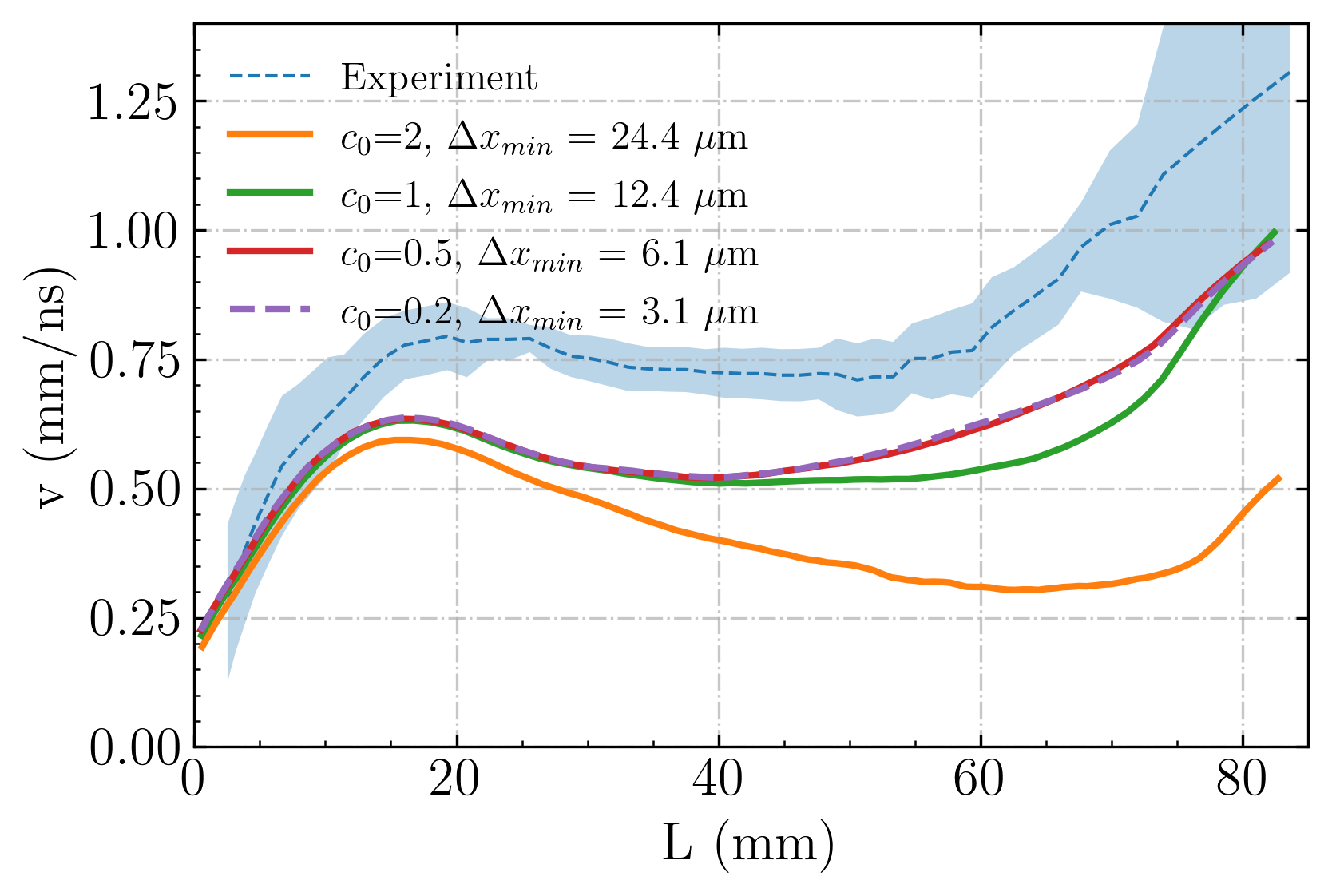}
	\caption{The streamer velocity versus the streamer length for streamers
          with different values for $c_0$ in the refinement criterion
          \mbox{$\Delta x < c_0 / \alpha (E)$}.
      }
	\label{fig:diff_mesh}
\end{figure}

Figure~\ref{fig:diff_mesh} shows the streamer velocity versus the streamer length for $c_0$ set to 2, 1, 0.5 and 0.2 and \mbox{$d_0\,=\,0.2 \,\mathrm{mm}$}. These parameters lead to a corresponding minimal grid spacing of 24.4, 12.2, 6.1 and 3.1\,$\mathrm{\mu m}$.
With \mbox{$c_0\,=\,2$} the grid is clearly too coarse and the streamer is much slower than for the other cases.
With \mbox{$c_0\,=\,1$} the results are similar to those on even finer grids, but the
streamer is a bit slower in the later stages. For \mbox{$c_0\,=\,0.5$} and \mbox{$c_0\,=\, 0.2$},
the streamer propagation is almost identical, indicating that the model is close to numerical convergence. For the results presented in this paper, we therefore use \mbox{$c_0 \,=\,0.5$ ($\Delta x_{min}$\,=\,6.1 $\mathrm{\mu m}$ )}.
Additionally, we also compared the effect of the parameter $d_0$, which controls the derefinement of the mesh. However, reducing $d_0$ to $10 \, \mathrm{\mu m}$ hardly affected the results, so we use \RV{\mbox{$d_0$\,=\,0.2\,mm}}.

\RV{For the simulations presented here, time integration was performed with Heun's method, a two-step explicit second order Runge-Kutta scheme,
for more details see section 2.4 of~\cite{teunissen2017}.}
\RV{The time step was limited according to}
\RV{
  \begin{eqnarray}
    \label{eq:dt-cfl}
    &\Delta t_\mathrm{CFL} \left(\frac{4 D_e}{\Delta x^2} + \sum \frac{v_i}{\Delta x} \right) \leq 0.5,\\
    &\Delta t_\mathrm{drt} \left(e \mu_e n_e / \varepsilon_0\right) \leq 1,\\
    &\Delta t = 0.9 \times \min(\Delta t_\mathrm{drt}, \Delta t_\mathrm{CFL}),
  \end{eqnarray}
}%
\RV{where $\Delta t_\mathrm{CFL}$ corresponds to a CFL condition (including diffusion), $\Delta t_\mathrm{drt}$ corresponds to the dielectric relaxation time, $\Delta t$ is the actual time step used in the simulations, and $\Delta x$ stands for $\Delta r$ or $\Delta z$, since they are equal. The simulations presented here are not sensitive to the time step, i.e., changing the safety factor from 0.9 to 0.5 hardly affects the results.}

For the case with \mbox{$c_0$\,=\,0.5}, the average time step for the streamer bridging
the gap was \mbox{$\Delta t\,=\,0.44 \, \textrm{ps}$}. \RV{Such a small time step was required due to a high electron density of about $10^{20}\,\textrm{cm}^{-3}$ occurring near the tip of the needle electrode.}
\RV{Typical cases took about 9 to 10 hours on a node with 24 Intel Xeon E5-2695\,v2\,@\,$2.4$\,GHz cores.}


\subsection{\RV{Effect of chemical reactions}}
\label{sec:chemical_reactions}

\RV{
A complete description of plasma-chemical reactions for non-equilibrium discharges in nitrogen-oxygen
mixtures is complex~\cite{kossyi1992}.
The chemical reactions considered in a fluid simulation can be as extensive as in~\cite{ono2020}, with hundreds of reactions, or as simple as in~\cite{bagheri2018}, considering only effective ionization rates.
Nine reactions are considered as the default in this paper, including three ionization reactions, two attachment reactions and four reactions related to light emission, as shown in table~\ref{tbl:reaction_table}.
To make clear how important different reactions are in our discharge regime, three other cases are investigated:
\begin{itemize}
	\item Case 1: reactions 1\,-\,3 and 6\,-\,9 from table~\ref{tbl:reaction_table}; ionization reactions and reactions related to light emission.
	\item Case 2: all the reactions from table~\ref{tbl:reaction_table} and reactions 10\,-\,11 from table~\ref{tbl:chemical_reactions}; adding two detachment reactions.
	\item Case 3: all the reactions from tables~\ref{tbl:reaction_table} and~\ref{tbl:chemical_reactions}; adding two negative ion conversion reactions (reactions 12\,-\,13), three positive ion conversion reactions (reactions 14\,-\,16), one electron-ion recombination reaction (reaction 17) and twelve ion-ion recombination reactions (reactions 18\,-\,29).
\end{itemize}
}

\begin{figure}[htb]
	\centering
	\includegraphics[width=0.9\linewidth]{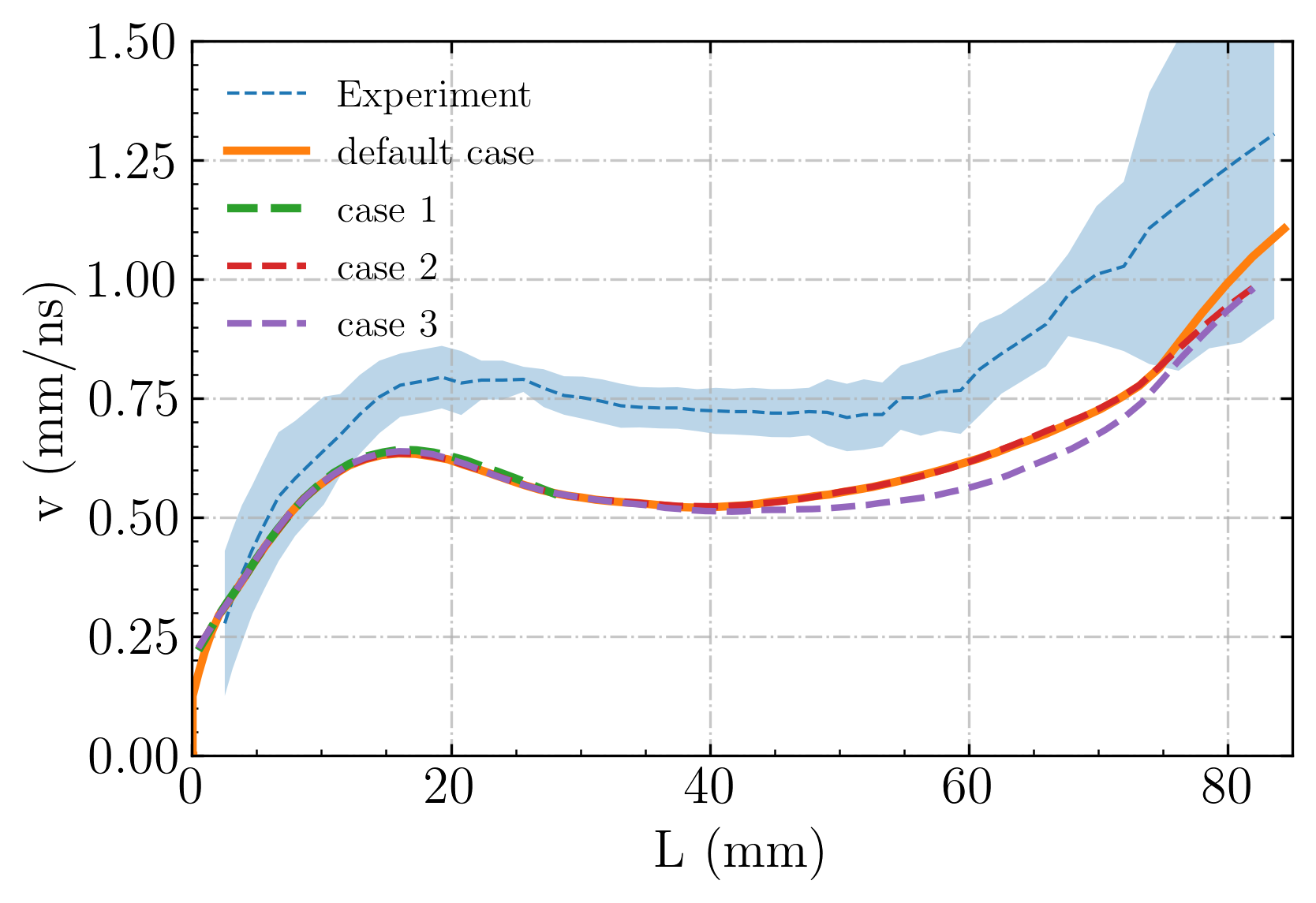}
	\caption{\RV{The streamer velocity versus the streamer length for streamers
			with different chemical reactions. The default case uses all the reactions from table~\ref{tbl:reaction_table};
			case 1, 2 and 3 are described in section~\ref{sec:chemical_reactions}.}}
	\label{fig:diff_reactions}
\end{figure}

\RV{
Figure~\ref{fig:diff_reactions} shows the streamer velocity versus the streamer length for the above three cases together with the default case. The results of all cases are similar. Note in particular how the inclusion of attachment and detachment reactions hardly makes a difference. The streamer is slightly slower for case 3, in which recombination is included, but this difference is much smaller than that between the experiment and the default case.
We therefore conclude that ionization reactions dominate the propagation of our discharge -- a streamer of $10^2$ ns time scale at 0.1\,bar with a background electric field of about 1.5\,kV/cm. Under these conditions, attachment, detachment and recombination appear to be less important for streamer propagation.
}

\begin{table*}
	\centering
	\caption{\RV{Additional chemical reactions. The effects of these reactions is studied in section \ref{sec:chemical_reactions}, see figure \ref{fig:diff_reactions}. Label \mbox{``u.s.''} stands for species which are not tracked in our simulation. $T_e$ in reaction rate $k_{17}$ is obtained from the mean electron energy ($\epsilon_e$) computed by Bolsig+ as \mbox{$T_e=2\epsilon_e/3k_B$}.}}
	\begin{tabular}{cccc}
		\hline
		Reaction No. & Reaction & Reaction rate coefficient  & Reference \\ \hline
		10 & $\textrm{O}_2^- + \textrm{M} \stackrel{k_{10}}{\longrightarrow} \textrm{e} + \textrm{O}_2 + \textrm{M}$ & $k_{10}=1.24\times10^{-17}\exp(-(\frac{179}{8.8+E/N})^2)\,\textrm{m}^3\textrm{s}^{-1}$ & \cite{pancheshnyi2013} \\
		11 & $\textrm{O}^- + \textrm{N}_2 \stackrel{k_{11}}{\longrightarrow} \textrm{e} + \textrm{NO}_2$ & $k_{11}=1.16\times10^{-18}\exp(-(\frac{48.9}{11+E/N})^2)\,\textrm{m}^3\textrm{s}^{-1}$ & \cite{pancheshnyi2013} \\
		12 & $\textrm{O}^- + \textrm{O}_2 \stackrel{k_{12}}{\longrightarrow} \textrm{O}_2^- + \textrm{O}$ & $k_{12}=6.96\times10^{-17}\exp(-(\frac{198}{5.6+E/N})^2)\,\textrm{m}^3\textrm{s}^{-1}$ & \cite{pancheshnyi2013} \\
		13 & $\textrm{O}^- + \textrm{O}_2 + \textrm{M} \stackrel{k_{13}}{\longrightarrow} \textrm{O}_3^- + \textrm{M}$ & $k_{13}=1.1\times10^{-42}\exp(-(\frac{E/N}{65})^2)\,\textrm{m}^3\textrm{s}^{-1}$ & \cite{pancheshnyi2013} \\
		14 & $\textrm{N}_2^+ + \textrm{N}_2 + \textrm{M} \stackrel{k_{14}}{\longrightarrow} \textrm{N}_4^+ + \textrm{M}$ & $k_{14}=5\times10^{-41}\,\textrm{m}^6\textrm{s}^{-1}$ & \cite{aleksandrov1999} \\
		15 & $\textrm{N}_4^+ + \textrm{O}_2 \stackrel{k_{15}}{\longrightarrow} \textrm{O}_2^+ + \textrm{N}_2 + \textrm{N}_2$ & $k_{15}=2.5\times10^{-16}\,\textrm{m}^3\textrm{s}^{-1}$ & \cite{aleksandrov1999} \\
		16 & $\textrm{O}_2^+ + \textrm{O}_2 + \textrm{M} \stackrel{k_{16}}{\longrightarrow} \textrm{O}_4^+ + \textrm{M}$ & $k_{16}=2.4\times10^{-42}\,\textrm{m}^6\textrm{s}^{-1}$ & \cite{aleksandrov1999} \\
		17 & $\textrm{e} + \textrm{O}_4^+ \stackrel{k_{17}}{\longrightarrow} \textrm{O}_2 + \textrm{O}_2$     & $k_{17}(E/N)=1.4\times10^{-12}(300\textrm{K}/T_e)^{1/2}\,\textrm{m}^3\textrm{s}^{-1}$ & \cite{kossyi1992} \\
		18 & $\textrm{N}_2^+ + \textrm{O}^- \stackrel{k_{18}}{\longrightarrow} \textrm{u.s.}$ & $k_{18}=10^{-13}\,\textrm{m}^3\textrm{s}^{-1}$ & \cite{kossyi1992} \\
		19 & $\textrm{N}_2^+ + \textrm{O}_3^- \stackrel{k_{19}}{\longrightarrow} \textrm{u.s.}$ & $k_{19}=10^{-13}\,\textrm{m}^3\textrm{s}^{-1}$ & \cite{kossyi1992} \\
		20 & $\textrm{N}_2^+ + \textrm{O}_2^- \stackrel{k_{20}}{\longrightarrow} \textrm{u.s.}$ & $k_{20}=10^{-13}\,\textrm{m}^3\textrm{s}^{-1}$ & \cite{kossyi1992} \\
		21 & $\textrm{O}_2^+ + \textrm{O}^- \stackrel{k_{21}}{\longrightarrow} \textrm{u.s.}$ & $k_{21}=10^{-13}\,\textrm{m}^3\textrm{s}^{-1}$ & \cite{kossyi1992} \\
		22 & $\textrm{O}_2^+ + \textrm{O}_3^- \stackrel{k_{22}}{\longrightarrow} \textrm{u.s.}$ & $k_{22}=10^{-13}\,\textrm{m}^3\textrm{s}^{-1}$ & \cite{kossyi1992} \\
		23 & $\textrm{O}_2^+ + \textrm{O}_2^- \stackrel{k_{23}}{\longrightarrow} \textrm{u.s.}$ & $k_{23}=10^{-13}\,\textrm{m}^3\textrm{s}^{-1}$ & \cite{kossyi1992} \\
		24 & $\textrm{O}_4^+ + \textrm{O}^- \stackrel{k_{24}}{\longrightarrow} \textrm{u.s.}$   & $k_{24}=10^{-13}\,\textrm{m}^3\textrm{s}^{-1}$ & \cite{kossyi1992} \\
		25 & $\textrm{O}_4^+ + \textrm{O}_2^- \stackrel{k_{25}}{\longrightarrow} \textrm{u.s.}$ & $k_{25}=10^{-13}\,\textrm{m}^3\textrm{s}^{-1}$ & \cite{kossyi1992} \\
		26 & $\textrm{O}_4^+ + \textrm{O}_3^- \stackrel{k_{26}}{\longrightarrow} \textrm{u.s.}$ & $k_{26}=10^{-13}\,\textrm{m}^3\textrm{s}^{-1}$ & \cite{kossyi1992} \\
		27 & $\textrm{N}_4^+ + \textrm{O}^- \stackrel{k_{27}}{\longrightarrow} \textrm{u.s.}$   & $k_{27}=10^{-13}\,\textrm{m}^3\textrm{s}^{-1}$ & \cite{kossyi1992} \\
		28 & $\textrm{N}_4^+ + \textrm{O}_2^- \stackrel{k_{28}}{\longrightarrow} \textrm{u.s.}$ & $k_{28}=10^{-13}\,\textrm{m}^3\textrm{s}^{-1}$ & \cite{kossyi1992} \\
		29 & $\textrm{N}_4^+ + \textrm{O}_3^- \stackrel{k_{29}}{\longrightarrow} \textrm{u.s.}$ & $k_{29}=10^{-13}\,\textrm{m}^3\textrm{s}^{-1}$ & \cite{kossyi1992} \\
		\hline
	\end{tabular}
	\label{tbl:chemical_reactions}
\end{table*}

\subsection{Transport data source}
\label{sec:transport_data}

Transport coefficients for fluid models can be computed from electron-neutral cross sections using
two-term or multi-term Boltzmann solvers~\cite{hagelaar2005, stephens2018, tejero-del-caz2019} or Monte Carlo swarm simulations~\cite{biagi1999, rabie2016}.
For N$_2$ and O$_2$, there are several sets of cross section available at LXCAT~\cite{Pancheshnyi_2012, carbone2021}. We here consider five such sets, namely those by Phelps~\cite{phelps1985, Phelps-database}, IST Lisbon~\cite{alves2014, loureiro1986, IST-Lisbon-database}, Morgan~\cite{Morgan-database}, TRINITI~\cite{TRINITI-database} and Biagi~\cite{biagi1999, Biagi-database}. 
It has been common practice to normalize and adjust the total cross sections so that the
transport coefficients computed with a Boltzmann solver agree well with
experimentally measured swarm data with isotropic scattering. For e.g. Phelps' cross sections, this was
done with a two-term method, whereas for Biagi's cross sections a Monte Carlo
method was used.
This means that even though multi-term and Monte Carlo methods are generally more
accurate than two-term approaches, they do not necessarily produce transport
coefficients that are closer to experimental data. In this section, we
investigate how different sets of cross sections and different Boltzmann solvers
affect transport coefficients and the agreement between our simulations and
experiments.

\begin{figure*}[htb]
	\centering
	\includegraphics[width=0.8\linewidth]{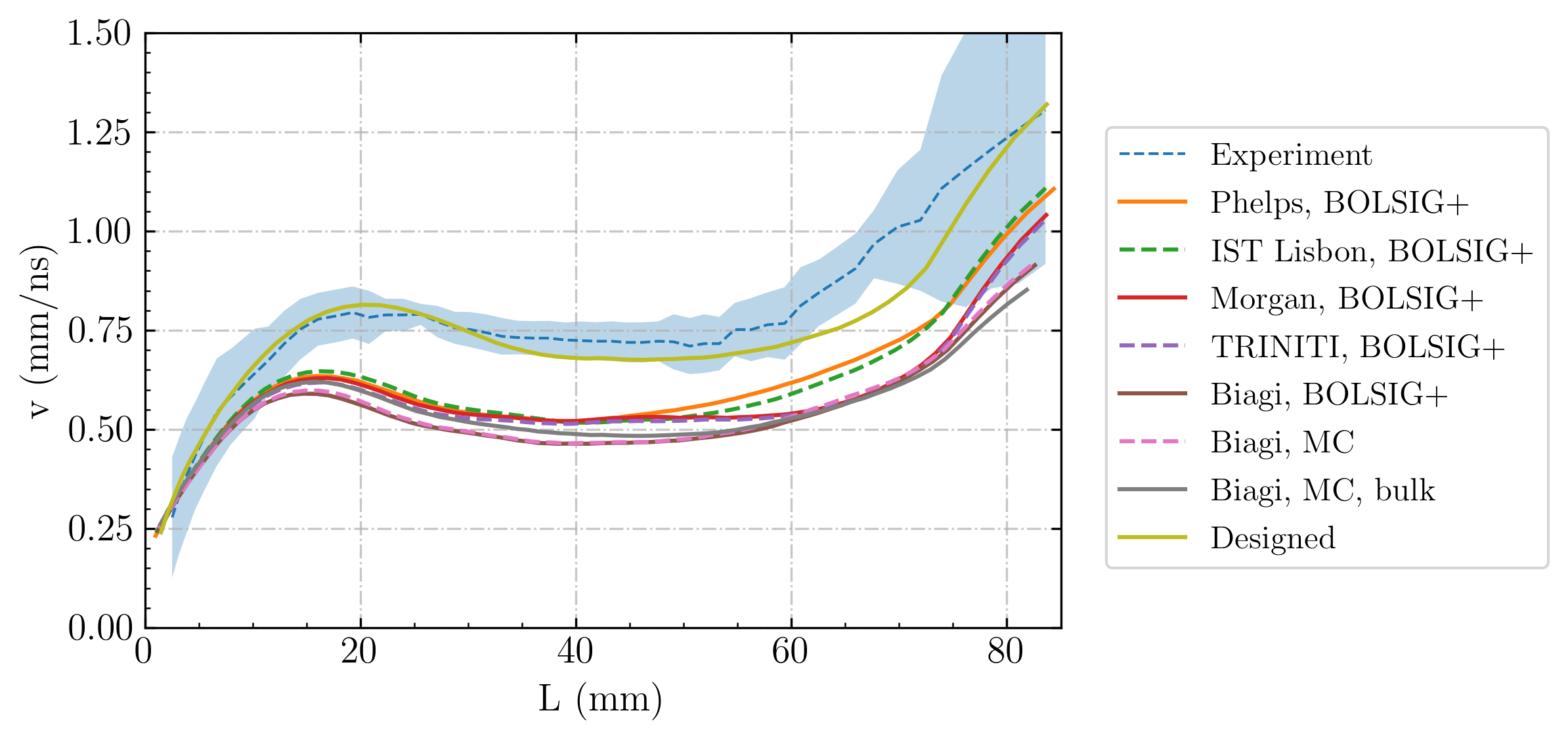}
	\caption{The streamer velocity versus the streamer length for
          simulations with different transport coefficients. The labels
          ``Phelps'', ``IST Lisbon'', ``Morgan'', ``TRINITI'' and ``Biagi'' indicate cross section databases, ``BOLSIG+'' and ``MC'' indicate the use of BOLSIG+ or a Monte Carlo Boltzmann solver, and ``bulk'' means that so-called bulk coefficients were used instead of flux coefficients.
          ``Designed'' is based on the ``Phelps, BOLSIG+'' database by increasing the ionization coefficient $\alpha$ and the mobility $\mu$ each with 20\%.}
	\label{fig:diff_cross_section}
\end{figure*}

We first used BOLSIG+~\cite{hagelaar2005} (a two-term Boltzmann solver) to calculate transport coefficients in 80\% N$_2$ and 20\% O$_2$ for the Phelps, IST Lisbon, Morgan, TRINITI and Biagi cross sections.
We used the online version BOLSIG+ via \url{lxcat.net}.
Figure~\ref{fig:diff_cross_section} shows how the streamer velocity in our simulations is affected by the resulting transport coefficients, which are shown in~\ref{sec:appendix-transport-coefficients}.
The streamers with the Phelps and IST Lisbon databases are fastest.
With Morgan and TRINITI data, the streamers are similar to those with Phelps data up to a length of $50 \, \textrm{mm}$, but thereafter they behave more like those with Biagi data.
The streamer with the Biagi database is the slowest, and it is about 10\% slower (at the same streamer length) than the fastest one. However, regardless of the cross sections used, all simulated velocities are significantly slower than the experimental one.

To investigate the influence of the type of Boltzmann solver we also computed transport data from Biagi's cross sections with a Monte Carlo code (available at \url{gitlab.com/MD-CWI-NL/particle_swarm}), which is similar to e.g.~\cite{rabie2016}. The resulting transport data is shown in~\ref{sec:appendix-transport-coefficients}.
With the Monte Carlo method, we computed both \emph{bulk} and \emph{flux} transport coefficients.
Bulk coefficients describe average properties of a group of electrons, taking ionization and attachment into account, whereas flux properties are averages for `individual’ electrons~\cite{dujko2013, petrovic2009}. 
The bulk mobility is larger than the flux one at high E/N because electrons that move faster than average also typically have higher energy, and hence produce more ionization.
The resulting streamer velocity with such Monte Carlo swarm flux and bulk data is shown in figure~\ref{fig:diff_cross_section}.
It can be seen that the choice of cross sections, Boltzmann solver and flux/bulk
coefficients does not significantly affect the streamer velocity, at least not
sufficiently to explain the observed discrepancy with the experimental results.

To match the experimental results, artificial transport coefficients were designed based on the Phelps database by increasing the ionization coefficient $\alpha$ and the mobility $\mu$ each with 20\%.
Figure~\ref{fig:diff_cross_section} shows that with these coefficients the relative error is often below 7\% when compared to the experimental velocity at the same length.


\subsection{Effect of background ionization density}
\label{sec:background-densities}

Positive streamers require free electrons ahead of them for their propagation, which can for example be provided by photoionization or background ionization.
Under the conditions considered here (air at 0.1\,bar, 50\,Hz repetition frequency), we generally expect photoionization to be the dominant source of free electrons. However, background ionization could play an important role in discharge inception~\cite{nijdam2011}. To investigate this, we have performed simulations with homogeneous background ionization densities of $10^{3}$, $10^{11}$, $10^{13}$ and $10^{15}$\,m$^{-3}$, in the form of electrons and positive ions. Photoionization was always included.

Note that a background ionization degree of $10^{3}$\,m$^{-3}$ corresponds to one electron per (10 cm)$^3$. In reality, having so few electrons would mean that inception would be unlikely within a $200 \, \textrm{ns}$ voltage pulse. Only electrons close to the electrode tip could start a discharge, since those farther away would quickly attach to oxygen molecules.
However, in a fluid model electrons are stored as densities which can lead to unrealistic streamer inception: an electron density in the zone above breakdown can represent a fraction of an electron, but it still can grow and rapidly start a discharge. Clearly, the continuum approximation of the fluid model breaks down in these cases. We nevertheless include this unrealistic case for demonstrative purposes.

\begin{figure}[htb]
	\centering
	\includegraphics[width=0.9\linewidth]{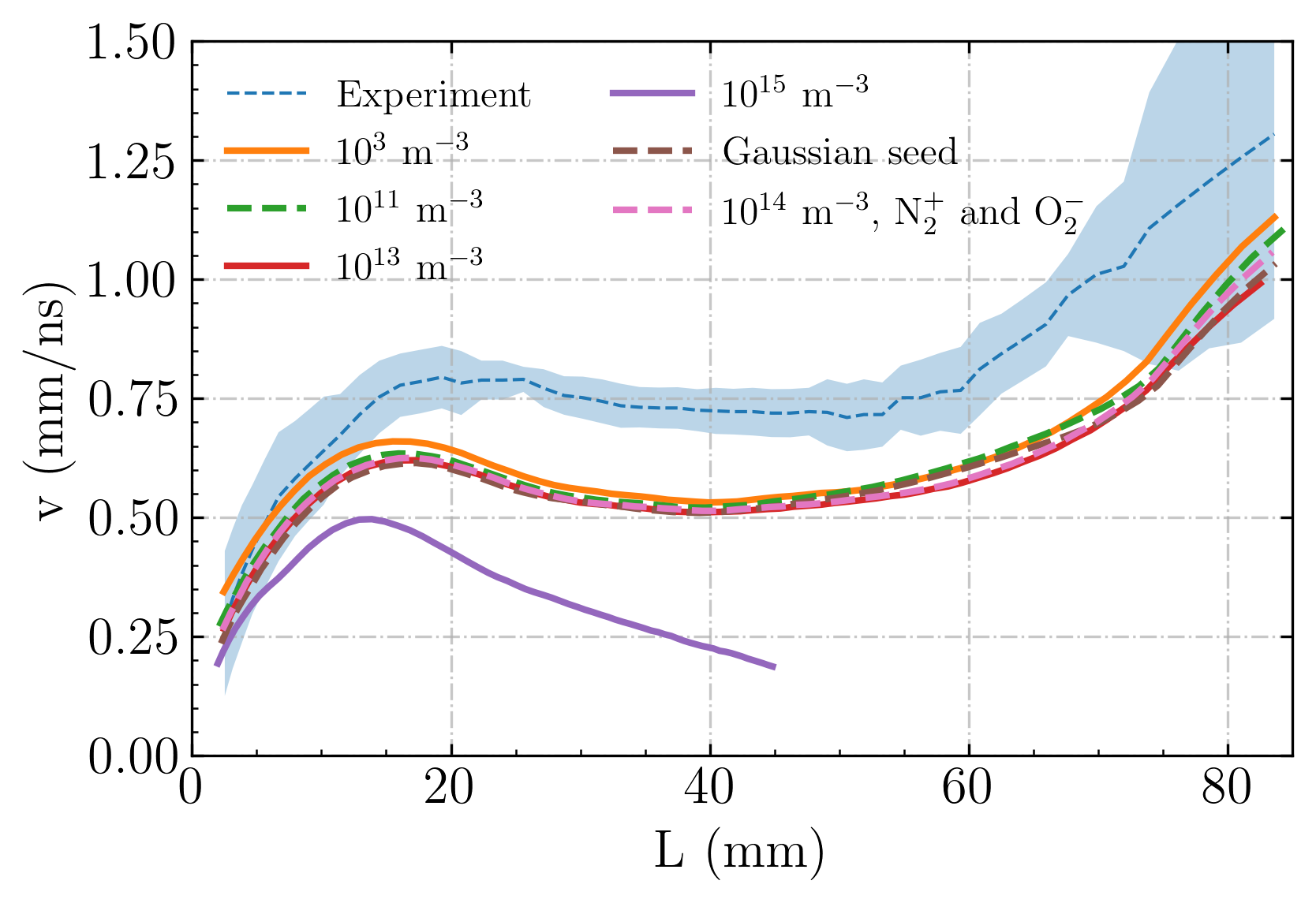}
	\caption{\RV{Streamer velocity versus streamer length for streamers
          with different uniform background ionization densities and a Gaussian
          initial seed. The curve labeled ``$10^{14}$\,m$^{-3}$, N$_2^+$ and O$_2^-$'' has a background ionization of $10^{14}$\,m$^{-3}$ N$_2^+$ and O$_2^-$. For the other curves the background species are electrons and N$_2^+$.
          In the rest of the paper, a uniform background
          ionization of $10^{11}$\,m$^{-3}$ electrons and N$_2^+$ is used.}
  }
	\label{fig:diff_background_ne}
\end{figure}

Figure~\ref{fig:diff_background_ne} shows that background ionization densities in the range of $10^{3}$ to $10^{13}$\,m$^{-3}$ have little effect on the streamer velocity versus streamer length.
With a lower background ionization degree the streamer starts a bit later, but there is no significant change in the velocity.
An even higher background ionization density of $10^{15}$\,m$^{-3}$ (corresponding to $10^{17}$\,m$^{-3}$ at 1\,bar) does lead to a significantly slower streamer. With this much background ionization the air surrounding the discharge has a non-negligible conductivity, reducing the field enhancement of the streamer.
Since expected background ionization levels under the conditions studied here are much lower, background ionization will probably not significantly affect the streamer velocity.

The cases discussed above included spatially uniform background ionization. To study the effect of more localized initial ionization, we have also performed simulations with a Gaussian initial seed located close to the tip of the needle electrode. A neutral seed consisting of electrons and positive ions was used, given by \mbox{$n_0\exp(-(d/R)^2)$}, with \RV{\mbox{$n_0\,=\,10^{14}$\,m$^{-3}$}}, $d$ the distance to the needle tip at \RV{\mbox{$(r, z) \,=\,(0\,\mathrm{mm},~90\,\mathrm{mm})$ }}and \mbox{$R\,=\,5\,\textrm{mm}$}. Besides this initial seed, no other (uniform) initial ionization was included.
The resulting streamer velocity is shown in figure~\ref{fig:diff_background_ne}.
The streamer velocity is almost the same as for the cases with a uniform background density of up to \RV{$10^{13}$\,m$^{-3}$}.

\RV{Remnants from previous pulses may affect the next streamer, in particular O$_2^-$ ions from which electrons can detach. We therefore include a case with a $10^{14}$\,m$^{-3}$ background density of positive (N$_2^+$) and negative ions (O$_2^-$) (see Fig. 2 of ~\cite{pancheshnyi2005}) and with two detachment reactions (reaction 10 and 11 from table~\ref{tbl:chemical_reactions}). The resulting streamer velocity versus length is similar to other cases with a background density of electrons and positive ions, as shown in figure~\ref{fig:diff_background_ne}.}

\RV{We conclude that some type of initial or background ionization is important for streamer inception, but that the stochastic nature of inception cannot be studied by a fluid model. The streamer propagation at later times is hardly affected by background ionization, at least under our conditions (air at 0.1\,bar), consistent with~\cite{nijdam2011, nijdam2014}.}


\subsection{Effect of the amount of photoionization}
\label{sec:photonionization}
As mentioned above, we expect photoionization to be the dominant source of free electrons ahead of the positive streamers studied here.
We now investigate how the amount
of photoionization affects streamer propagation.
We adjust the amount of photoionization by changing the proportionality factor $\xi$ in equation~(\ref{eq:photoi-source}).
Four cases are considered: \mbox{$\xi = 0.075$}, as is used in the rest of this paper - this value is taken from~\cite{zheleznyak1982} considering the electric field at our streamer head - and \mbox{$\xi$ = 0.05}, 0.0075 and 0.75.

\begin{figure}[htb]
	\centering
	\includegraphics[width=0.9\linewidth]{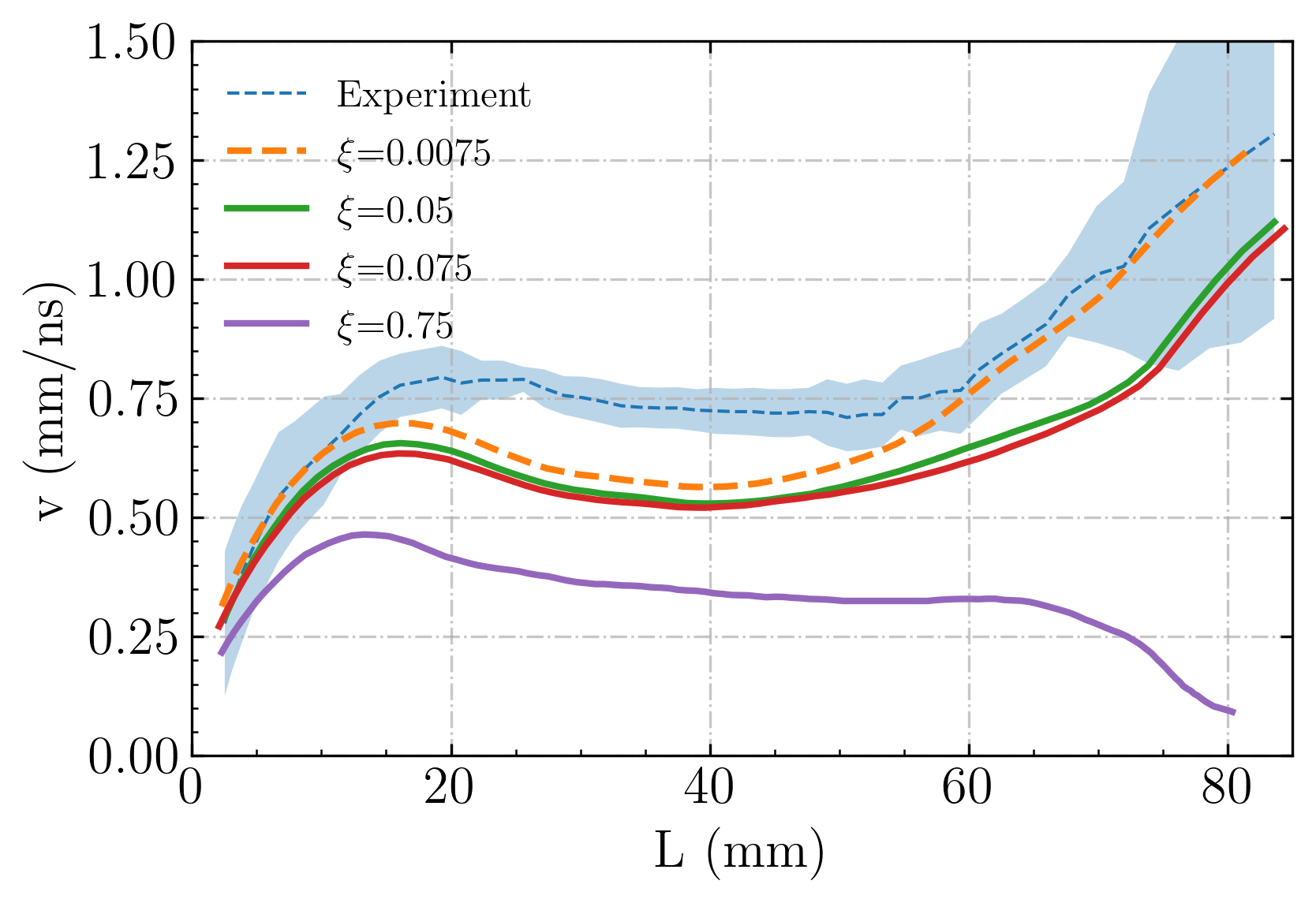}
	\caption{The streamer velocity versus the streamer length for streamers with different amounts of photoionization, see equation (\ref{eq:photoi-source}). The value of $\xi$ used in the rest of the paper is $0.075$.}
	\label{fig:photoionization}
\end{figure}


Figure~\ref{fig:photoionization} shows the streamer velocity versus streamer length for these four cases, using a uniform background density of $10^{11}$\,m$^{-3}$.
With ten times less photoionization, the streamer velocity increases at later times\RV{, approaching the experimental streamer velocity. This behavior, which at first seems surprising, shows the nonlinear nature of streamer discharges. Less photoionization leads to sharper electron density gradients at the streamer head, a smaller radius, and a higher degree of ionization, which can result in a higher electric field and a higher streamer velocity. However, note that there is still a qualitative discrepancy between the results of this case (\mbox{$\xi$ = 0.0075}) and the experimental velocity in the range of \mbox{15\,mm < L < 50\,mm}.}
With ten times more photoionization, the streamer is significantly slower. The electron density around the streamer head then increases sufficiently to reduce its field enhancement, as also happened in section \ref{sec:background-densities} with a high background ionization density of $10^{15} \, \textrm{m}^{-3}$.
However, if the amount of photoionization is only slightly changed using \mbox{$\xi = 0.05$} (the smallest tabulated value in~\cite{zheleznyak1982}), the streamer velocity is hardly affected, as shown in figure~\ref{fig:photoionization}.

Additionally, we have also repeated the above simulations with an even lower background ionization density, but the results were almost identical. This indicates that even if photoionization is reduced by a factor ten, it still dominates over a background density of $10^{11}$\,m$^{-3}$. Finally, note that all results were obtained at a pressure of $0.1$\,bar, at which there is less quenching than at 1\,bar, see section~\ref{sec:sim-model}.


\subsection{Effect of gas temperature}
\label{sec:gas-temperature}

\begin{figure}[htb]
	\centering
	\includegraphics[width=0.9\linewidth]{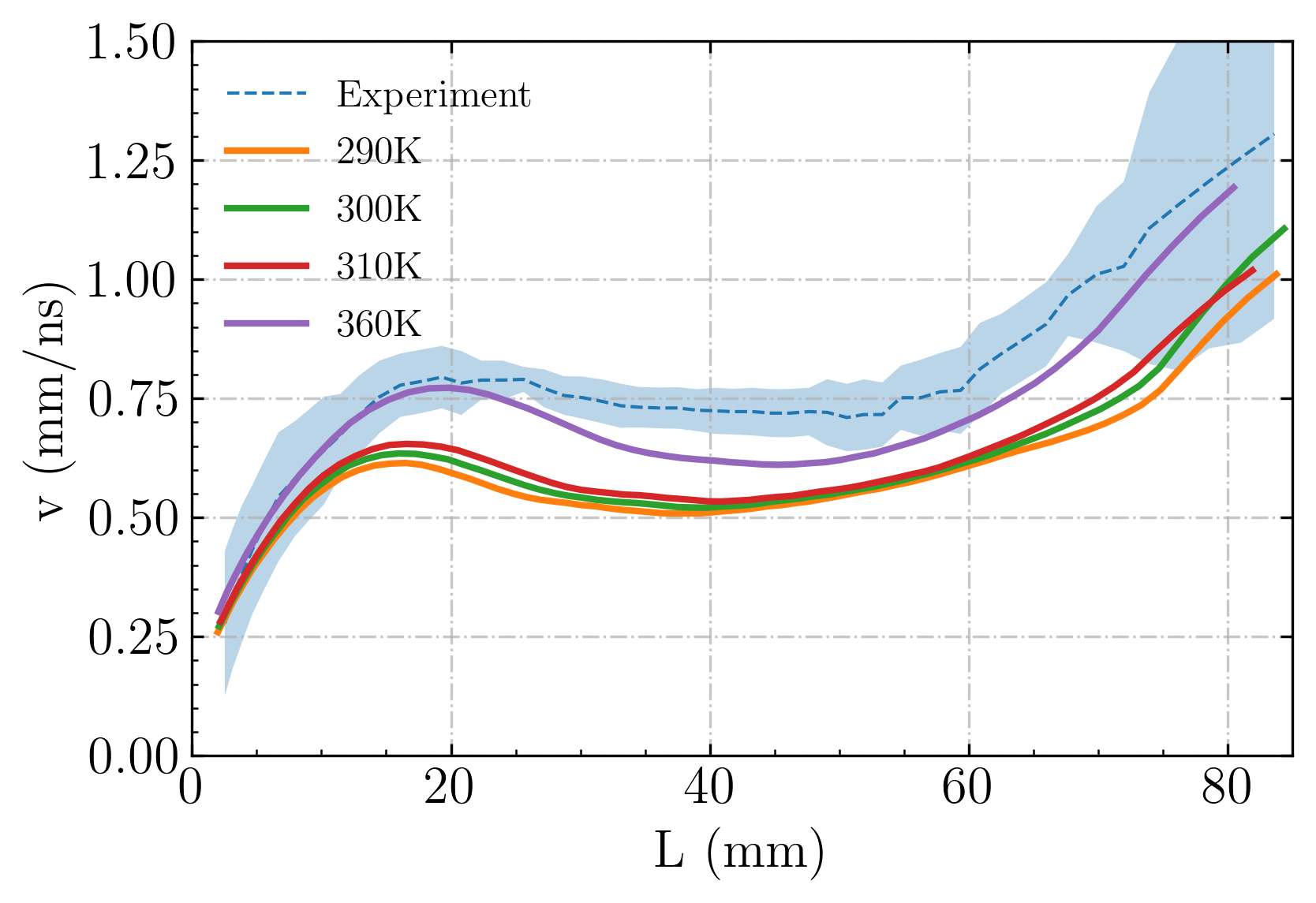}
	\caption{The streamer velocity versus the streamer length at different gas temperatures. In the rest of this paper, a gas temperature of 300\,K is used.}
	\label{fig:diff_gas_temp}
\end{figure}

In our experiments the lab temperature was about 293\,K, but the gas temperature in the vessel was not directly measured, and we have thus far assumed it to be 300\,K.
The two main factors affecting the gas temperature in the vessel are heating due to repetitive discharges and cooling due to the expansion of the compressed artificial air flowing into the vessel. To investigate the effect of temperature variations, we have performed simulations with gas temperatures of 290\,K, 300\,K, \RV{310\,K} and 360\,K. \RV{The gas pressure is always 0.1\,bar in the simulations.}
Figure~\ref{fig:diff_gas_temp} shows the streamer velocity for these four cases, together with the experimental result. The average streamer velocity between the two electrodes is 0.58, 0.59, 0.60 and 0.69\,mm/ns for the cases at 290\,K, 300\,K, \RV{310\,K} and 360\,K, respectively. For a 10\,K change in the gas temperature, the change in the streamer velocity at the same length is about 3\%, on average. When the gas temperature increases 20\% to 360\,K, the simulated streamer velocity is closer to the experimental data, and the velocity error at the same length is less than 15\%.

That a higher gas temperature leads to a higher streamer velocity is to be expected, because it leads to a higher value of $E/N$ in the discharge gap, just as when the applied voltage is increased. 
In our model, the gas number density is computed using the ideal gas law, so a reduction in gas pressure has a similar effect as an increase in temperature. However, \RV{the gas pressure was controlled to be 0.1\,bar in the experiments, with an uncertainty of about 1\%, so a change in pressure} cannot account for the observed discrepancies.

We can roughly estimate the temperature increase caused by the repetitive discharges. From the voltage-current waveform, we estimate that about $2 \, \textrm{mJ}$ is deposited in the plasma per 200\,ns pulse. At 50\,Hz repetition frequency, this corresponds to \mbox{$P = 0.1$\,W} of heating power. The gas flush rate in the experiments was \mbox{$f = 2 \, \textrm{SLM} \approx 2 \times 10^1 \, \textrm{L/min}$}. Dry air at 0.1\,bar and 300\,K has a specific heat capacity \mbox{$C_p = 1.0 \, \textrm{kJ/(kg K)}$}, a density \mbox{$\rho = 0.12 \, \textrm{kg/m}^3$} and a thermal diffusivity \mbox{$\alpha = 2.2 \times 10^{-4} \, \mathrm{m}^2\mathrm{/s}$}. If we assume heating happens uniformly and neglect losses to the vessel walls, then a rough estimate for the temperature increase would be \mbox{$\Delta T = P / (C_p \rho f) \approx 2 \, \mathrm{K}$}.
Alternatively, we could assume that heat is predominantly produced in the axial streamer channel and that heat diffusion occurs only in the radial direction. This results in an `effective' volume of order \mbox{$\pi \alpha h t$}, where \mbox{$h = 10 \, \textrm{cm}$} is the gap size and $t$ the time. The temperature increase in this volume can then be estimated as \mbox{$\Delta T = P / (C_p \rho \pi \alpha h) \approx 1 \times 10^1 \, \mathrm{K}$}. This is a rough estimate, not accounting for e.g., wall losses or the actual flow pattern in the vessel, nor the fact that the temperature close to the center could be considerably higher. We only have preliminary experimental data on the temperature increase, obtained with Raman scattering and optical emission spectroscopy. These measurements indicated a $\Delta T$ in the range of $10^1 \, \textrm{K}$ to $10^2 \, \textrm{K}$, consistent with the estimate given above. We therefore conclude that gas heating might explain part of the observed differences between simulations and experiments.




\subsection{Effect of applied voltage}
\label{sec:voltage-effect}

\begin{figure}[htb]
	\centering
	\includegraphics[width=0.9\linewidth]{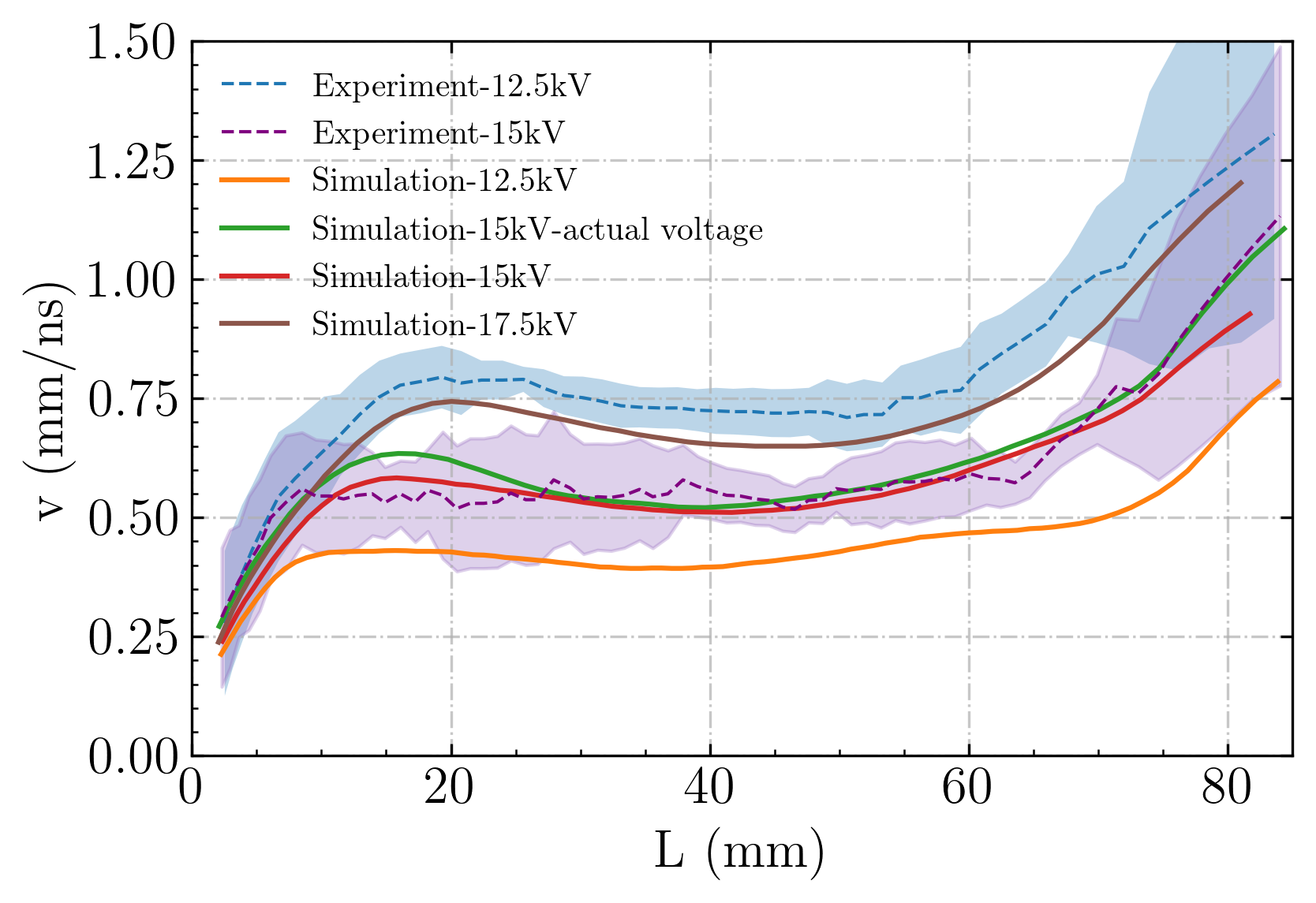}
	\caption{Streamer velocity versus streamer length in simulations and experiments at different applied voltages. Note that the experimental results at 12.5\,kV show larger fluctuations than those at 15\,kV. This happens because each frame is taken from a new streamer, and discharge inception at lower voltages is more stochastic.
	}
	\label{fig:12.5kv}
\end{figure}

\RV{The uncertainty in the measured applied voltage is only about 2\%, which is unlikely to account for the observed discrepancies in streamer velocity. However, out of scientific curiosity, we nevertheless investigate the effects of the voltage amplitude and rise time on streamer propagation below.}
Figure~\ref{fig:12.5kv} shows the streamer velocity in simulations at 12.5\,kV, 15\,kV and 17.5\,kV, together with
experimental data at 12.5\,kV and 15\,kV.
Note that for the curve labeled ``Simulation-15kV-actual voltage'', the voltage is applied according to the actual waveform used in the experiment, as shown in figure~\ref{fig:voltage_time}. For the other cases, the applied voltage rises linearly from zero to the maximum voltage within 65\,ns, after which it is constant.
The streamer evolution is similar for the cases with an actual voltage waveform and the linearly-rising 15\,kV voltage waveform, but the streamer is a little bit faster with the actual waveform, since it has a slight overshoot.
In all cases, the velocity profiles follow the same pattern: the velocity first increases, then it decreases slightly, and finally it increases again as the streamers approach the opposite electrode. As expected, streamer velocities increase for higher applied voltages. The simulated streamers are always slower than the experimental ones at the same applied voltage. On average, the velocity in a simulation at 17.5\,kV agrees quite well with the experimental velocity at 15\,kV.
However, since the experimental uncertainty in the voltage is only about 2\%, this cannot explain the observed discrepancies.

\begin{figure}[htb]
	\centering
	\includegraphics[width=0.9\linewidth]{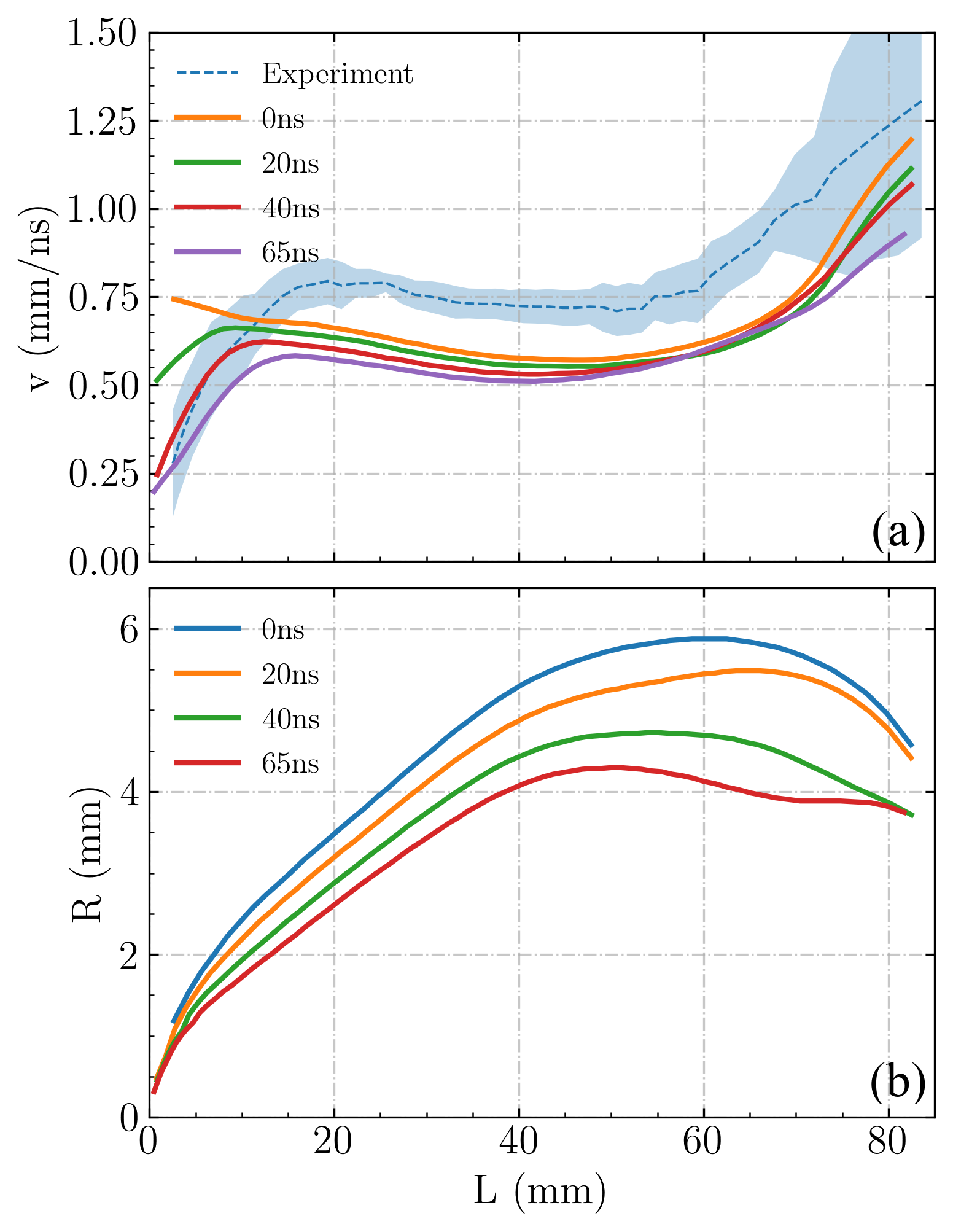}
	\caption{
		The streamer velocity (a) and streamer radius (b) versus the streamer length for streamers with different voltage rise times.}
	\label{fig:diff_risetime}
\end{figure}

We have also studied the effect of the voltage rise time on streamer propagation,
using an applied voltage of 15\,kV and a variable linear voltage rise.
Figure~\ref{fig:diff_risetime} shows the streamer velocity and streamer radius versus the streamer length for voltage rise times of 0, 20, 40 and 65\,ns.
\RV{The streamer with 0\,ns rise time starts immediately when the simulation begins.
Inception, here identified by a reduction in the maximal electric field, takes longer with a longer voltage rise time.
With a rise time of 20\,ns, 40\,ns and 65\,ns, the streamers incept at 10\,ns, 20\,ns and 30\,ns, respectively.}
With a shorter rise time, the streamer velocity is initially higher.
As the streamers get longer they propagate at the applied voltage and velocity differences become smaller when compared at the same length. Because the voltage rise time has an effect on the conductivity of the initial part of the streamer channel, small differences in velocity remain, with slightly higher velocities for shorter rise times. \RV{When comparing the velocity at the same streamer length (\mbox{30\,mm < L < 80\,mm}), the streamer velocity (averaged over length) of the 0\,ns rise time case increases by about 10\% compared to the 65\,ns case.} But there is still about a 20\% discrepancy compared to the experiments.
That a faster voltage rise leads to a higher streamer velocity was also found in~\cite{Komuro_2013}.
As in~\cite{Komuro_2013}, we also observe a larger streamer radius with a shorter voltage rise time, see figure~\ref{fig:diff_risetime}~(b).

A related effect is that with a shorter voltage rise time, the electric field initially exceeds the breakdown threshold in a larger area around the needle electrode.
This leads to a wider and more conductive streamer channel connected to the electrode. At later discharge stages the internal electric field in this part of the channel can therefore be lower while carrying the same electric current, which lead to less light emission around the tip of the electrode.

\subsection{Finite plate electrode vs infinite plate electrode}
\label{sec:finite-plate-electrode}

In this paper, we apply a potential profile at the upper and lower domain boundaries to make the simulations consistent with the experimental electrode geometry, as described in section \ref{sec:computational-domain}. This potential profile depends in particular on the radius of the HV electrode in which the needle is embedded, see figure~\ref{fig:simulation-domain}. If this electrode has a small radius, then the voltage will drop more rapidly in its vicinity, leading to a background field that is higher close to the electrode and lower farther away from it. If both the grounded and HV electrodes instead have a very large radius the voltage drop will be approximately linear, and the background field homogeneous.

\begin{figure}[htb]
	\centering
	\includegraphics[width=0.9\linewidth]{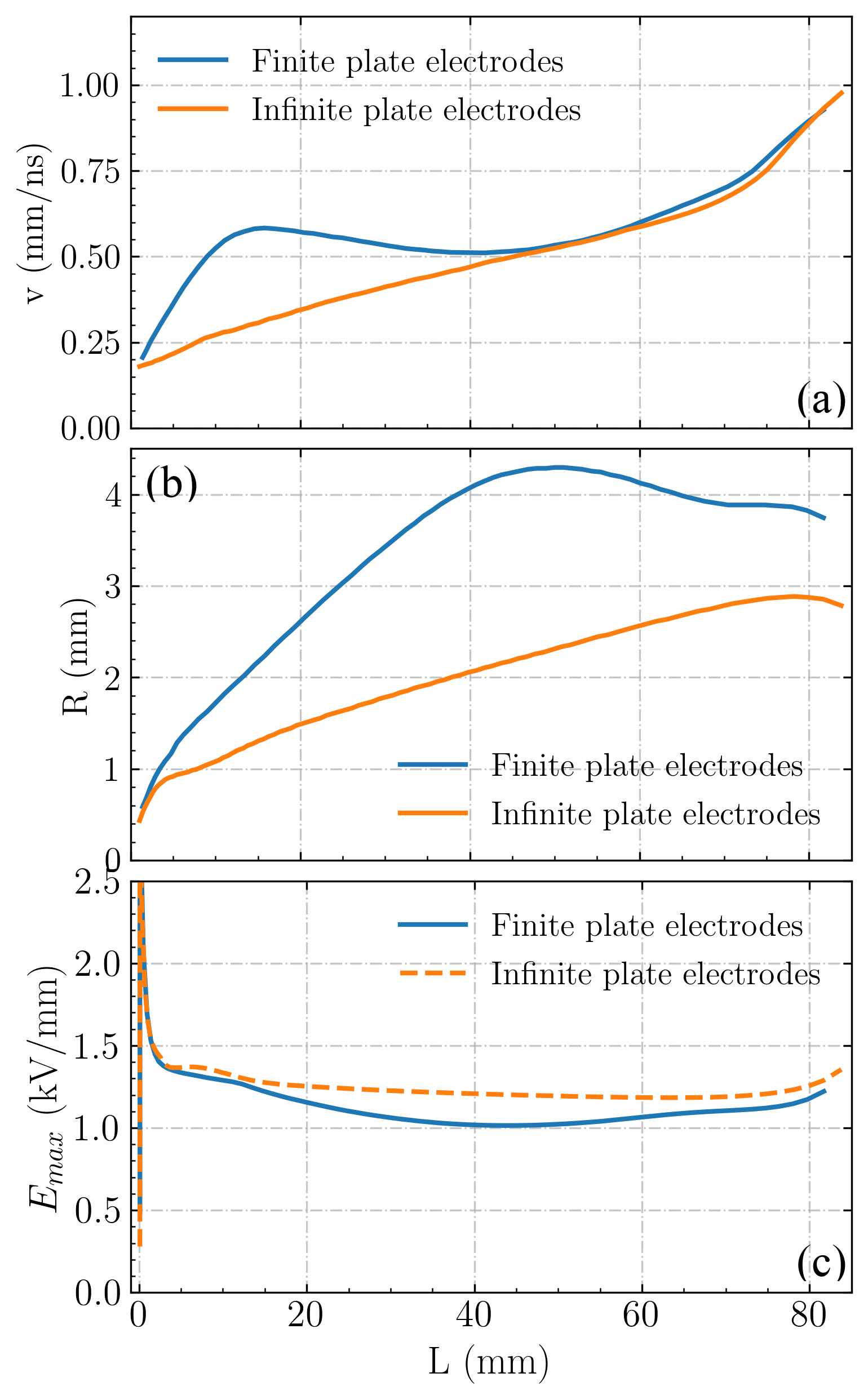}
	\includegraphics[width=0.98\linewidth]{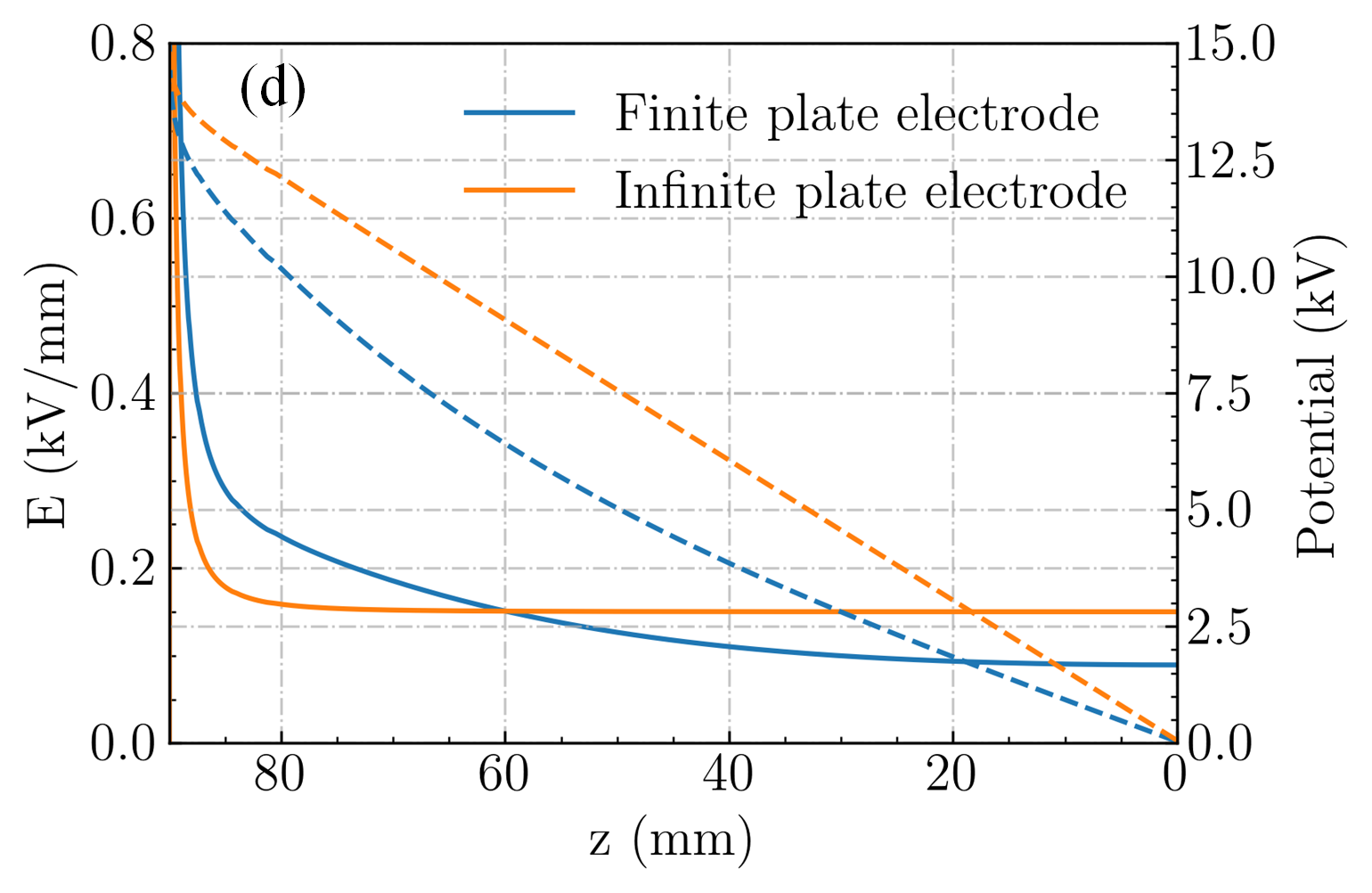}

	\caption{ The streamer velocity (a), the streamer radius (b) and the maximum electric field (c) versus the streamer length for streamers with finite plate electrodes and infinite plate electrodes, both with a needle electrode protruding 10\,mm into a 10\,cm wide gap. (d) The electric field and potential distribution along the z axis in the absence of space charge.
	The solid lines are for electric fields, and the dashed lines for electric potentials.}
	\label{fig:finite-vs-infinite}
\end{figure}

Here, we compare simulation results for the experimental electrode geometry with results using quasi-infinite plate electrodes and the same 10\,mm long protruding needle electrode. These `infinite' electrodes are incorporated by applying a voltage uniformly on the upper and lower domain boundaries.
We use a linearly increasing voltage with a rise time of 65\,ns for both cases.
Figure~\ref{fig:finite-vs-infinite} (a), (b) and (c) show the streamer velocity, streamer radius, and the maximal electric field at the streamer head for these two cases. The background electric field and the potential along the z axis for these two cases are shown in figure~\ref{fig:finite-vs-infinite} (d).
The use of infinite plate electrodes leads to a couple of clear differences:
\begin{itemize}
  \item The voltage drop between the electrodes is now approximately linear at \mbox{0\,mm < z < 85\,mm}, whereas with finite electrodes this drop is steeper near the HV electrode.
  \item The streamer velocity increases approximately linearly with streamer length, in contrast to the pattern of acceleration, deceleration and acceleration with finite electrodes.
  \item The streamer velocity is initially significantly lower, but when streamers have nearly bridged the whole gap their velocities are similar regardless of electrode geometry.
  \item The maximal electric field at the streamer head is now almost constant between 5\,mm and 75\,mm, whereas a decrease and consecutive increase are visible with finite electrodes.
  \item The background electric field is almost constant in the area \mbox{0\,mm < z < 85\,mm}, whereas it continuously decreases from the needle electrode to the ground with finite electrodes.
  \item The streamer is thinner than with finite electrodes, and the streamer radius keeps increasing until the streamer length is about 80\,mm.
\end{itemize}

\subsection{\RV{Effect of boundary conditions}}
\label{sec:boundary_conditions}

\begin{figure}[htb]
	\centering
	\includegraphics[width=0.9\linewidth]{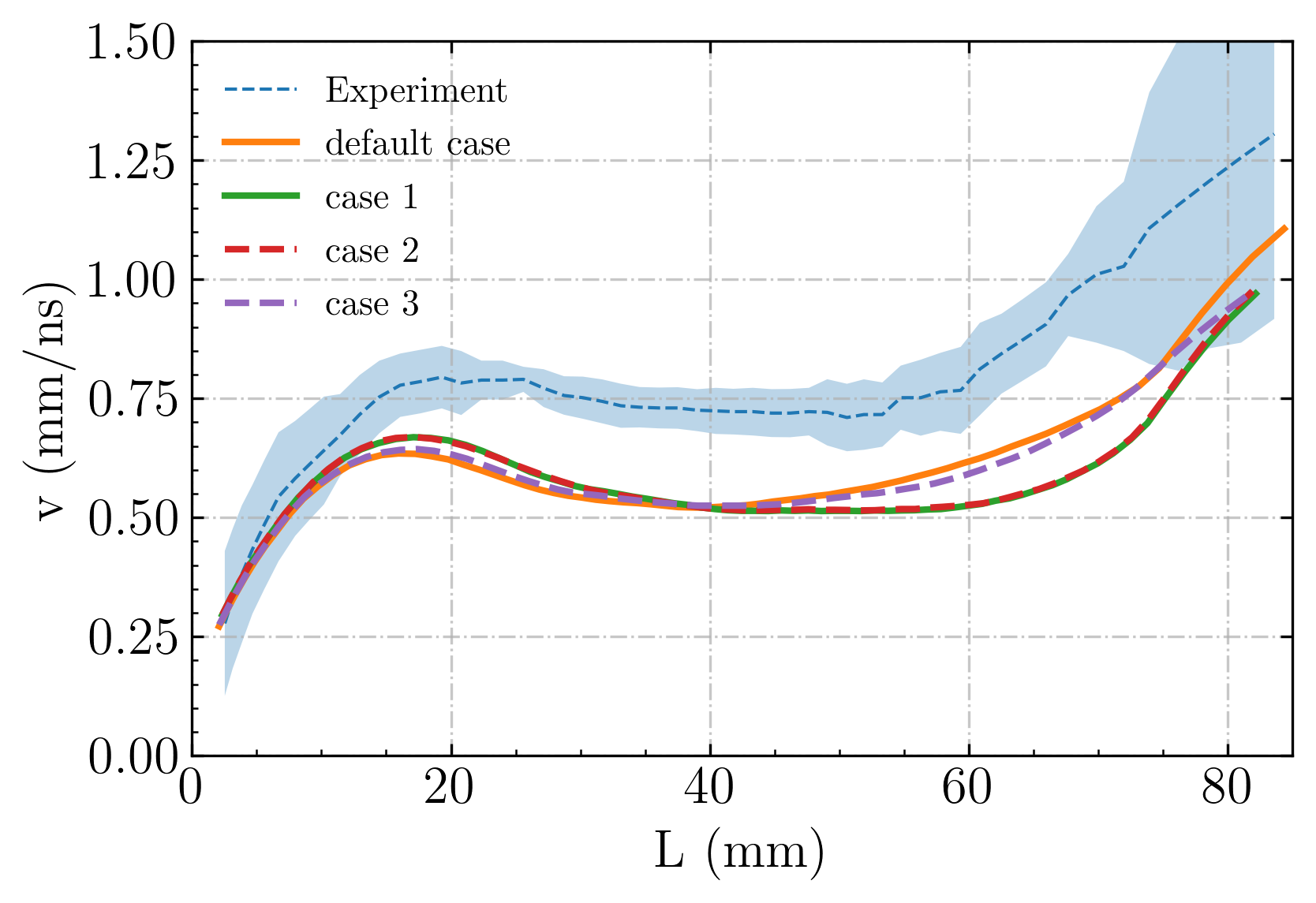}
	\caption{\RV{Streamer velocity versus streamer length for different boundary conditions for the electric potential. The default case uses homogeneous Neumann boundary conditions in the radial direction, whereas case 1, 2 and 3 use Dirichlet boundary conditions, see section \ref{sec:boundary_conditions} for details.}}
	\label{fig:diff_bc}
\end{figure}

\RV{The experiments are performed in a \emph{quasi}-cylindrical vessel with a diameter of 32.4\,cm and a height of 38.0\,cm. The simulation domain does not capture the whole vessel, see figure \ref{fig:simulation-domain}, so electrostatic boundary conditions for the simulation domain need to be carefully set.
  The upper and lower boundaries use pre-computed Dirichlet boundary conditions considering the effect of the finite plate electrodes, as described in section~\ref{sec:computational-domain}.
  	For our default case, these values were precomputed with a FEM method for a fully axisymmetric discharge vessel with a 16\,cm radius, in the absence of a discharge.
	However, the discharge vessel contains observation windows and gas in and outlets, so it is not fully axisymmetric, as shown in~\ref{sec:vessel}.
	In particular, it contains a large window of 10\,cm radius located 26\,cm away from its center.
        Furthermore, we have thus far applied homogeneous Neumann boundary conditions for the potential at the radial boundary, whereas some type of Dirichlet boundary condition might be more appropriate.
        To investigate the effect of these boundary conditions for the potential, we compare our default case with three other cases:}
\RV{
	\begin{itemize}
		\item Case~1: Identical to the default case, but using the FEM solution as a Dirichlet boundary condition on the radial boundary (instead of homogeneous Neumann).
		\item Case~2: A larger \mbox{16\,cm $\times$ 10\,cm} computational domain, now using a Dirichlet zero boundary condition on the radial boundary.
		\item Case~3: Identical to case 1, but now the electric potential was pre-computed for a larger discharge vessel with a radius of 26\,cm. This larger radius could account, to some extent, for the windows it contains.
	\end{itemize}
}
\RV{Figure~\ref{fig:diff_bc} shows the streamer velocity versus length for all cases.
  Case~1 and and case~2 give similar results. Compared to the default case, streamer velocities are first slightly higher, but in the range \mbox{50\,mm < L < 80\,mm} they are lower. This implies that the use of radial Dirichlet boundary conditions reduces the potential at the streamer head at later stages. On the one hand, the agreement between case~1 and case~2 shows that our computational domain is sufficiently large for these cases, so that the discharge and boundary conditions are only weakly coupled. On the other hand, the disagreement with the default case indicates that this coupling is significantly stronger with homogeneous Neumann boundary conditions.
  If we instead use boundary conditions pre-computed for a larger discharge vessel (case 3), the streamer velocity is similar to the default case.}

\RV{It is difficult to say which of these cases most closely matches the
  experiments, as the actual discharge charge vessel is not axisymmetric and
  contains windows. That these windows play a role was confirmed experimentally,
  because the streamers propagated slightly off-axis, with the deviation towards
  the largest window. The default case and case 3 seem to give slightly better
  qualitative agreement in the streamer velocity, but this could just be
  coincidence. However, what we can conclude is that our results are sensitive
  to the used electrostatic boundary conditions. For future validation studies,
  this could mean there is a trade-off in the size of windows for optical access: large windows
  facilitate measurements, but they make it harder to accurately model the
  electrostatic boundary conditions.}

\subsection{Other findings}

Discharges around the edge of the HV plate electrode were observed in both the experiments and simulations. In the simulations we suppress these discharges by artificially reducing the ionization coefficient around the edge of the plate electrode to zero.

\section{Summary}
\label{sec:summary}

We have quantitatively compared simulations and experiments of single positive streamers in artificial air at $0.1 \, \textrm{bar}$.
Good qualitative agreement is observed between the experimental and simulated optical emission profiles. In both cases, the streamers have similarly shaped bright heads, and darker tails. The streamer velocity and radius also show good qualitative agreement. After inception, the streamers first accelerate, then they slowly decelerate, and finally they accelerate again when approaching the grounded electrode. Quantitatively, the simulated streamer velocity is about 20\% to 30\% lower at the same streamer length, and the simulated radius is about 1\,mm (20\% to 30\%) smaller.
These discrepancies could be explained by a temperature increase in the
experiments due to 50\,Hz repetitive pulses.

\paragraph{Possible errors in the experimental measurements}
\begin{itemize}
  \item In the experiments, only preliminary measurements were available for gas
  temperature variations in the vessel. These indicate that the gas temperature
  due to previous discharges could locally rise by roughly 10 to 100\,K. A
  temperature increase towards the upper end of this range could explain much of
  the observed differences between simulations and experiments.
  \item There are fluctuations in the streamer velocity obtained from the
  experimental images, since each image corresponds to a different discharge.
  The experimental velocity therefore has an intrinsic error of about 10\% when
  compared at a particular position or time.
  \item The experimental uncertainty in the applied voltage is about 2\%, and in
  the gas pressure it is about 1\%. The observed discrepancies in streamer
  velocity can therefore not be explained by errors in these parameters.
\end{itemize}

\paragraph{Possible errors in the simulations}
\begin{itemize}
  \item The streamer properties in a fluid simulation depend on the used
  transport coefficients. The cross-section databases used
  here~\cite{phelps1985, loureiro1986, biagi1999, alves2014} are often based on
  data obtained decades ago. It is difficult for us to assess the accuracy and
  uncertainty in this data, but more having more recent cross-section data would
  be helpful for the validation of simulation models.
  \item We have used a fluid model with the local field approximation. Previous
  studies have shown that the predictions of this model can deviate from those
  of particle-in-cell simulations, see e.g.~\cite{markosyan2015}. However, based
  on recent unpublished comparisons of axisymmetric particle and fluid models in
  our group, we think such model error is unlikely to account for the
  observed discrepancies.
  \item Related to the above point, discharge inception can sometimes not
  accurately be modeled with a fluid model, since the continuum approximation
  breaks down when there are few particles. This could perhaps also account for
  some of the observed discrepancies.
  \item \RV{The experiments were performed with a 50 Hz repetition rate, but the simulations did not take into account remnants from previous pulses. An accumulation of long-lived excited species could for example lead to increased ionization rates. However, our simulations have been proved to be quite insensitive to initial ionization conditions.}
  \item \RV{The experimental vessel contains several windows, of which one is large, and it is not fully axisymmetric. This leads to uncertainty in the boundary conditions for the electric potential in the simulations. Our simulation results are sensitive to these boundary conditions. Experimentally, an off-axis devation of the streamer towards the largest window was observed.}
\end{itemize}

\paragraph{Summary of results for parameter studies}



\begin{enumerate}
  \item \RV{The propagation of the discharge considered here -- a streamer developing on $10^2$ ns time scale at 0.1\,bar with a background electric field of about 1.5\,kV/cm -- is mostly controlled by ionization reactions. Attachment, detachment and recombination reactions have a much smaller effect.}
  \item Using transport coefficients computed from different cross-section databases affects the simulated streamer evolution. However, the simulated velocities are always significantly lower than those in the experiments. The choice of Boltzmann solver (BOLSIG+ or Monte Carlo particle swarms) has little effect on the velocity. By artificially increasing both the ionization coefficient and the mobility by 20\%, the simulated streamer velocity is much closer to the experimental one.
  \item Increasing the applied voltage increases streamer velocities. With a 17.5\,kV applied voltage, the simulated streamer velocity is similar to the experimental one at 15\,kV. However, the experimental uncertainty in the voltage is only about 2\%. A longer voltage rise time initially slows down the streamers, but its effect is weaker at later times and longer streamer lengths.
  \item Initial or background ionization is essential for streamer inception, but it hardly affects streamer propagation at later times. However, a very high background ionization level leads to slower streamers, as it reduces the field enhancement at their heads.
  \item With ten times less photoionization, the streamer velocity increases by up to 30\%, in particular when the streamer has almost bridged the gap. However, the velocity profile then differs qualitatively from the experimental measurements. With ten times more photoionization, streamers are significantly slower, as they lose some of their field enhancement due to the relatively high degree of ionization ahead of them. 
  \item A higher gas temperature leads to higher E/N values. For a 10\,K change in the gas temperature, the change in the streamer velocity at the same length is about 3\%. When the gas temperature is 360\,K in the simulations, the difference between the simulated and experimental streamer velocity is less than 15\%.
  \item The size of the plate electrodes changes the background electric field in the gap. This affects the maximum electric field at the streamer head, and can lead to qualitatively different streamer propagation between the electrodes.
  With quasi-infinite plate electrodes, the streamer velocity monotonically increases within the gap.
\end{enumerate}

\section*{Availability of model and data}

The source code and documentation for the model used in this paper are available at \url{gitlab.com/MD-CWI-NL/afivo-streamer} (git commit \RV{\texttt{872d3827}}) and at \url{teunissen.net/afivo_streamer}. A snapshot of the code, data and experimental images is available at \RV{\url{doi.org/10.5281/zenodo.4905873}}.

\section*{Acknowledgments}

X.L. and S.D. were supported by STW-project 15052 ``Let CO$_2$ Spark''. X.L. was also supported by the National Natural Science Foundation of China (51777164).\\
\RV{We would like to thank both anonymous referees for their valuable suggestions.}

\section*{PhD student contributions}


X.L.~performed the simulations and comparisons. S.D.~built the experimental setup, performed the experiments and contributed to the data analysis.


\appendix

\section{\RV{The discharge vessel}}
\label{sec:vessel}

\begin{figure}[htb]
	\centering
	\includegraphics[width=0.9\linewidth]{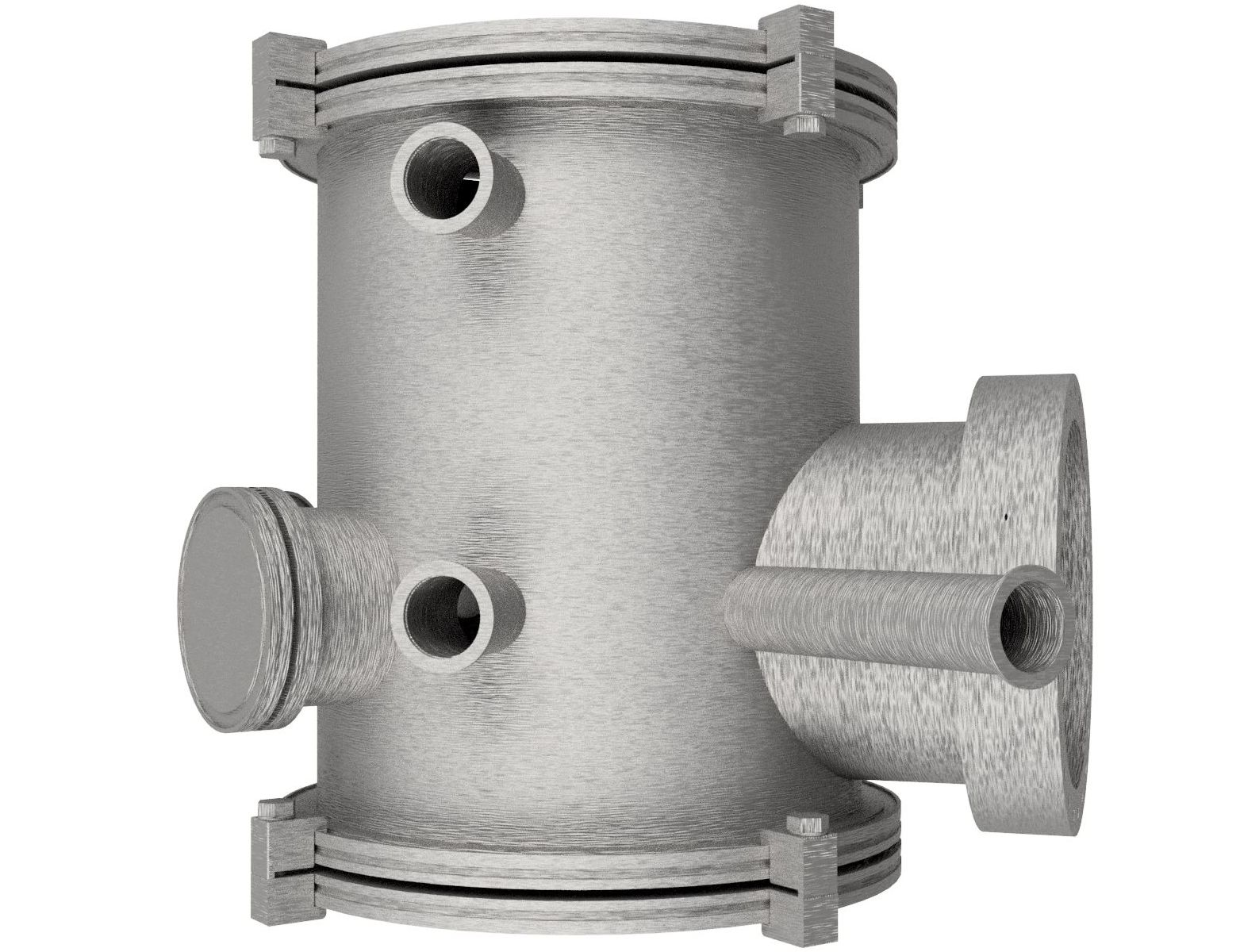}
	\caption{\RV{The geometry of the experimental discharge vessel, as seen
            from the side. It is a quasi-cylindrical vessel that contains
            observation windows and gas flow tubes. The ICCD camera captures
            pictures through large window on the right. We observed that
            streamers propagated slightly off-axis, with the deviation towards
            the largest window, which is most likely related to the electric
            potential distribution inside the vessel, see section
            \ref{sec:boundary_conditions}.}}
	\label{fig:vessel}
\end{figure}

\RV{Figure~\ref{fig:vessel} shows the geometry of the experimental discharge vessel.}

\section{Data acquisition methods from experimental images}
\label{sec:appendix-a}

\begin{figure}[htb]
	\centering
	\includegraphics[width=1.0\linewidth]{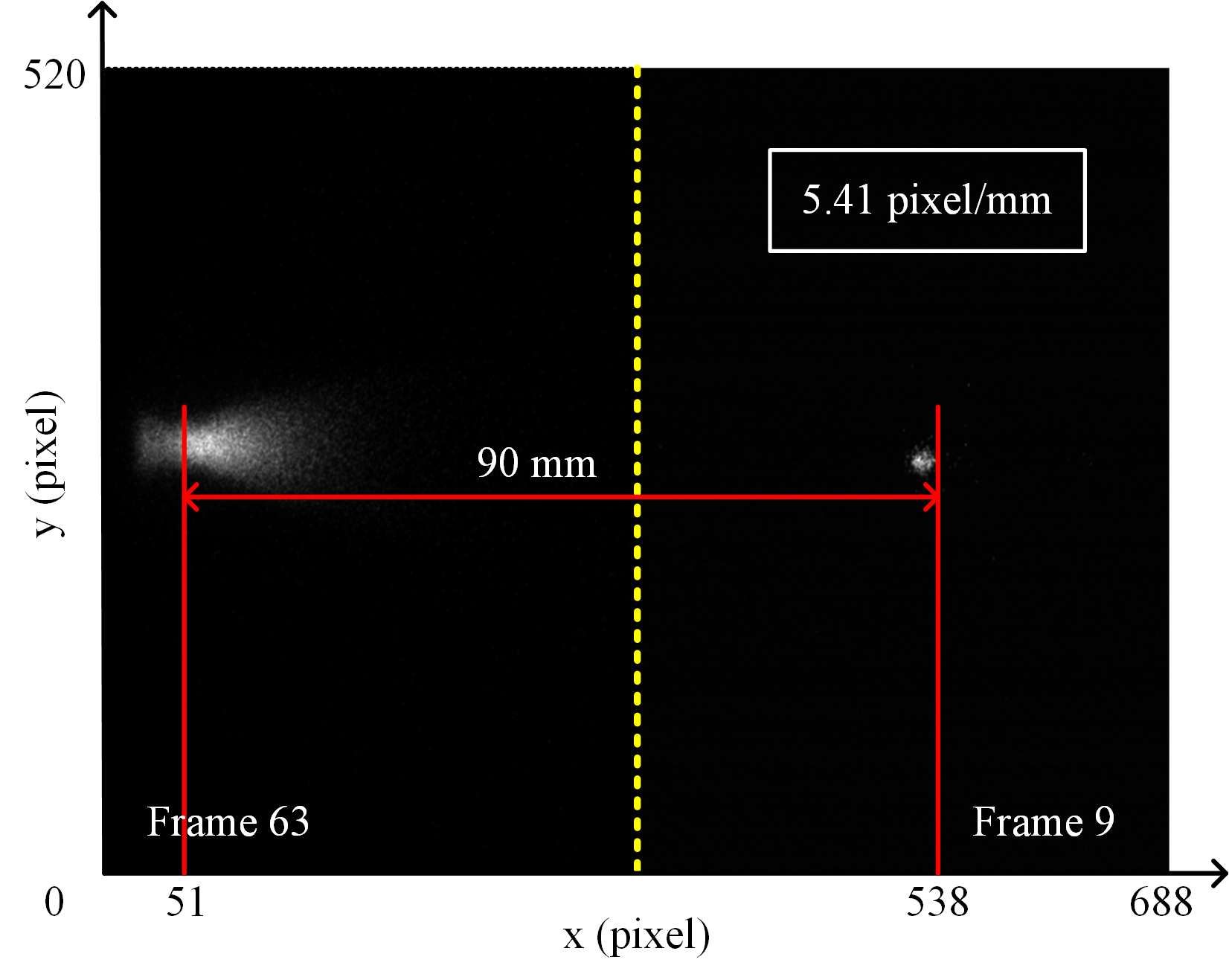}
	\caption{Illustration showing how distances were determined from the
          experimental images. Left: first frame in which the streamer touches
          the grounded electrode. A reflection is visible. Right: first frame in
          which bright emission is visible near the HV electrode tip.}
	\label{fig:raw-images}
\end{figure}

Here we explain how distances were determined from the experimental images, using
figure~\ref{fig:raw-images}. The size of the experimental images is \mbox{688 $\times$ 520}
pixels. The right side of figure~\ref{fig:raw-images}
shows 
the first frame in which bright emission can be observed. Assuming that light
first appears around the tip of the needle electrode, the tip is located around
x-pixel\,538. The left side of figure~\ref{fig:raw-images}
shows 
the first frame in which the streamer touches the grounded electrode. The light
reflected by the flat electrode indicates that it is located around x-pixel\,51.
The distance between the needle electrode tip and the grounded electrode is 90\,mm, so on this image 5.41\,pixels correspond to 1\,mm. This conversion factor is then used to determine the streamer length and radius in all the images.

\section{Transport coefficients value from different sources}
\label{sec:appendix-transport-coefficients}

\begin{figure*}[htb]
	\centering
	\includegraphics[width=0.8\linewidth]{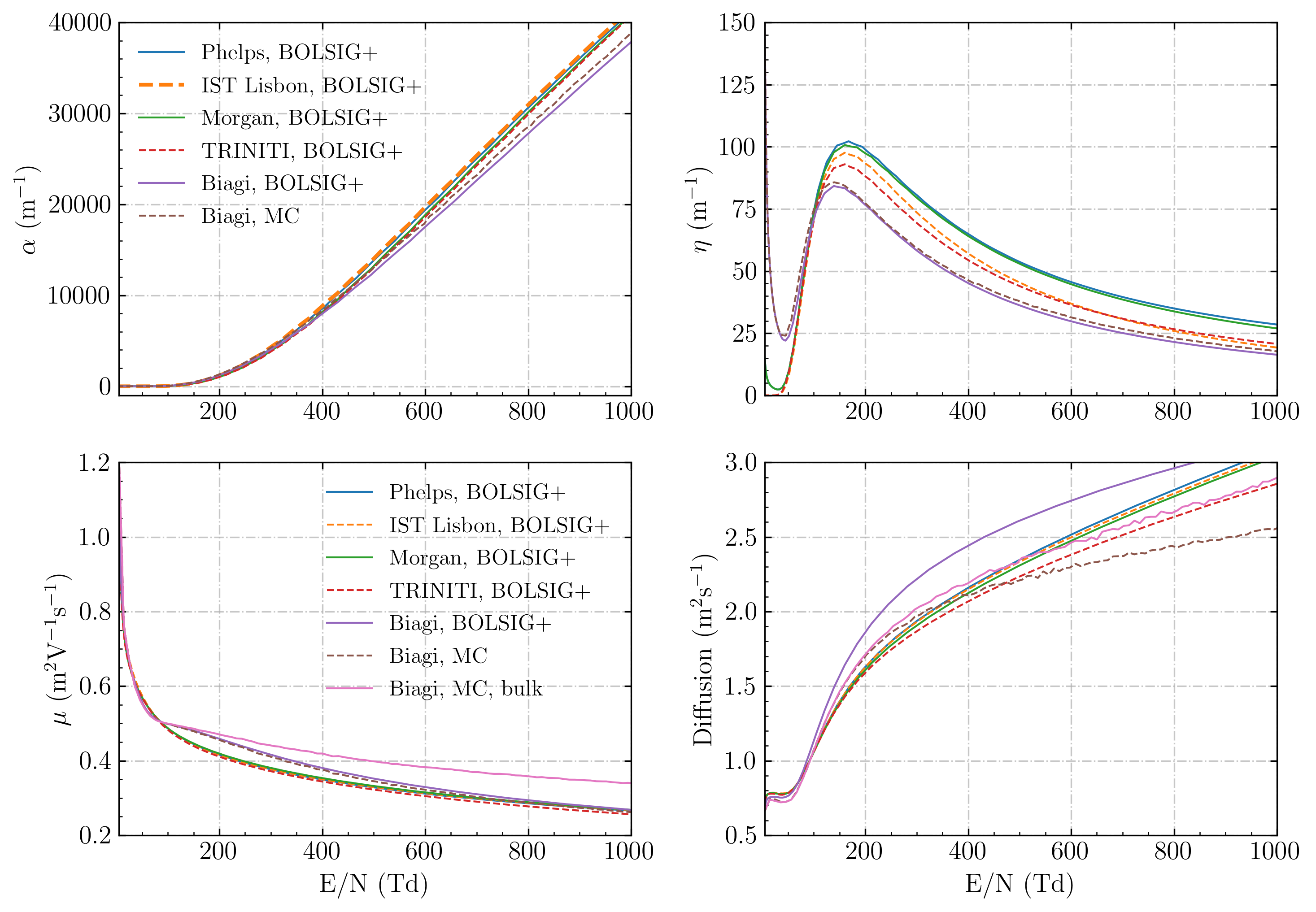}
	\caption{Transport coefficients ($\alpha, \eta, \mu$ and Diffusion) as determined from several sets of cross sections and different Boltzmann solvers.
          The same labels are used as in figure~\ref{fig:diff_cross_section}.
        }
	\label{fig:trasport_value}
\end{figure*}

Figure~\ref{fig:trasport_value} gives the ionization ($\alpha$), attachment ($\eta$), mobility ($\mu$) and diffusion coefficients used in section~\ref{sec:transport_data}.


\section{Needle electrode vs initial ionized seed}
\label{sec:electrode-vs-seed}

In previous computational studies, an elongated ionized seed with an equal density 
of electrons and positive ions was often used as a pseudo-electrode to start a streamer, see e.g.~\cite{bagheri2020a, xiaoran-psst-2020}.
Due to electron drift such a seed becomes electrically screened, leading to a high electric field at its tip that can start a streamer discharge, depending on the shape and density of the seed~\cite{luque2008}.
However, to quantitatively compare simulations with experiments, we have here instead implemented an actual needle electrode in our field solver. This ensures that the electric potential at the electrode contour is equal to the applied voltage.

\begin{figure}[htb]
	\centering
	\includegraphics[width=0.9\linewidth]{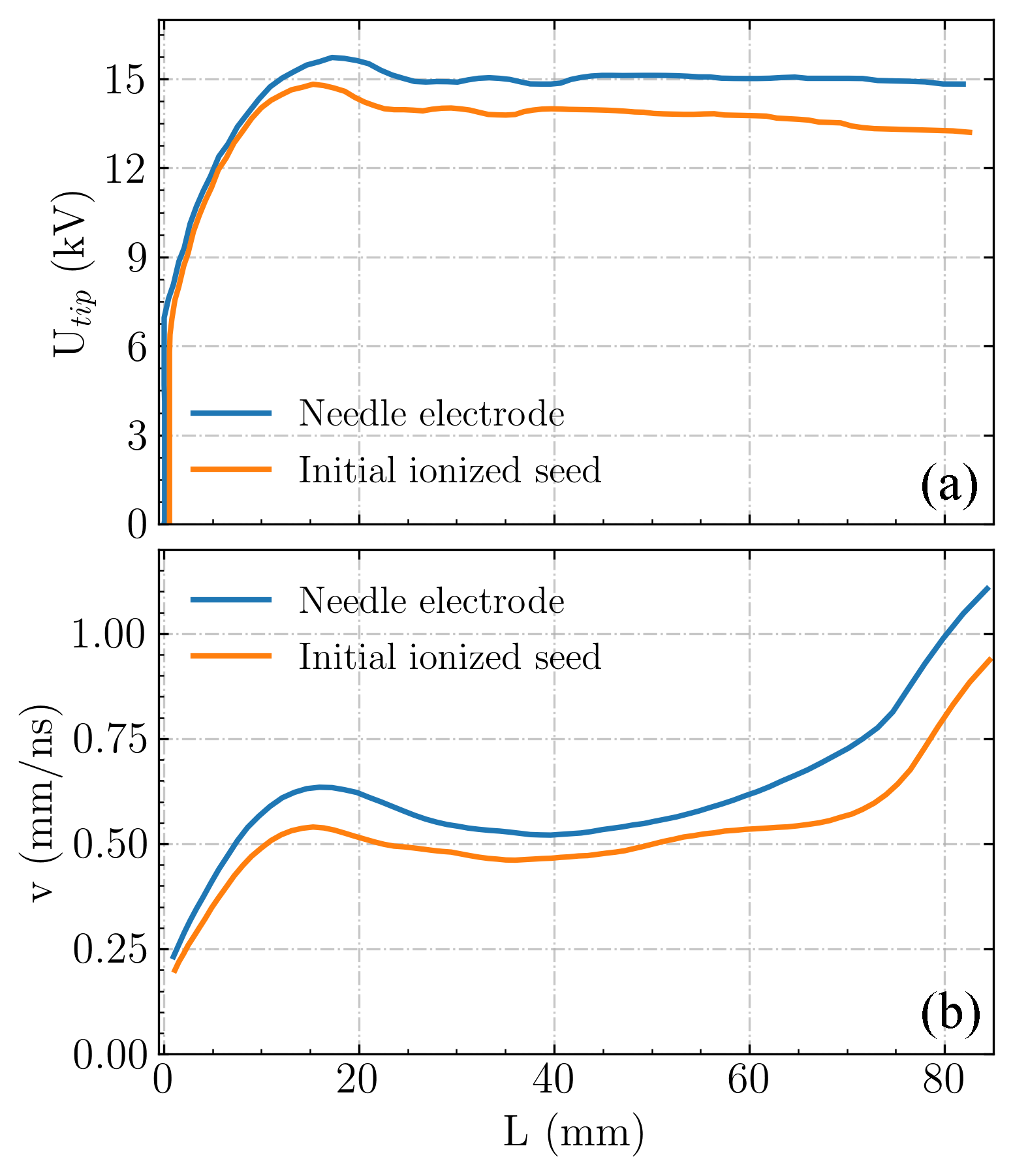}
	\caption{Comparison of streamers originating from an ionized seed (see
          text) and a needle electrode. (a) Electric potential at the
          seed/needle electrode lower tip. (b) Streamer velocity versus streamer
          length.}
	\label{fig:electrode-vs-seed}
\end{figure}

To compare streamers originating from a needle electrode to those originating from an ionized seed, we ran a simulation with an ionized seed of about 10\,mm long with a radius of about 0.5\,mm.
The electron and $\textrm{N}_2^+$ density were $10^{19}$\,m$^{-3}$ at its center, with a decay at a distance above \mbox{$d = 0.3 \, \textrm{mm}$} using a so-called smoothstep profile: \mbox{$1 - 3 x^2 + 2 x^3$} up to \mbox{$x = 1$}, where
\mbox{$x = (d - 0.3 \, \mathrm{mm})/0.3 \mathrm{mm}$}~\cite{teunissen2017}.
Figure~\ref{fig:electrode-vs-seed}~(a) shows the evolution of the electric potential at the tip of the seed and the needle electrode.
With an actual electrode, the potential at the needle tip agrees with the applied voltage (shown in figure~\ref{fig:voltage_time}).
But with an ionized seed, the actual potential at the seed tip is lower due to the seed's finite conductivity, and it essentially becomes part of the streamer. In other words, there is a potential drop between the plate electrode and the tip of the former seed. The streamer originating from an ionized seed is therefore slower, as shown in figure~\ref{fig:electrode-vs-seed}~(b).

\section*{References}

\bibliography{references_zotero}

\providecommand{\newblock}{}
\begin{thebibliography}{10}
\expandafter\ifx\csname url\endcsname\relax
  \def\url#1{{\tt #1}}\fi
\expandafter\ifx\csname urlprefix\endcsname\relax\def\urlprefix{URL }\fi
\providecommand{\eprint}[2][]{\url{#2}}

\bibitem{ebert2010}
Ebert U, Nijdam S, Li C, Luque A, Briels T and van Veldhuizen E 2010 {\em
  Journal of Geophysical Research: Space Physics\/} {\bf 115} ISSN 2156-2202

\bibitem{laroussi2014}
Laroussi M 2014 {\em Plasma Processes and Polymers\/} {\bf 11} 1138--1141 ISSN
  1612-8869

\bibitem{bardos2010}
B{\'a}rdos L and Bar{\'a}nkov{\'a} H 2010 {\em Thin Solid Films\/} {\bf 518}
  6705--6713 ISSN 0040-6090

\bibitem{popov2016}
Popov N~A 2016 {\em Plasma Sources Science and Technology\/} {\bf 25} 043002
  ISSN 0963-0252

\bibitem{bruggeman2016}
Bruggeman P~J, Kushner M~J, Locke B~R, Gardeniers J~G~E, Graham W~G, Graves
  D~B, {Hofman-Caris} R~C~H~M, Maric D, Reid J~P, Ceriani E, Rivas D~F, Foster
  J~E, Garrick S~C, Gorbanev Y, Hamaguchi S, Iza F, Jablonowski H, Klimova E,
  Kolb J, Krcma F, Lukes P, Machala Z, Marinov I, Mariotti D, Thagard S~M,
  Minakata D, Neyts E~C, Pawlat J, Petrovic Z~L, Pflieger R, Reuter S, Schram
  D~C, Schr{\"o}ter S, Shiraiwa M, Tarabov{\'a} B, Tsai P~A, Verlet J~R~R, von
  Woedtke T, Wilson K~R, Yasui K and Zvereva G 2016 {\em Plasma Sources Science
  and Technology\/} {\bf 25} 053002 ISSN 0963-0252

\bibitem{nijdam2020}
Nijdam S, Teunissen J and Ebert U 2020 {\em Plasma Sources Science and
  Technology\/} {\bf 29} 103001 ISSN 1361-6595

\bibitem{babaeva2021}
Babaeva N~Y and Naidis G~V 2021 {\em Journal of Physics D: Applied Physics\/}
  {\bf 54} 223002 ISSN 0022-3727, 1361-6463

\bibitem{marskar2020}
Marskar R 2020 {\em Plasma Sources Science and Technology\/} {\bf 29} 055007
  ISSN 1361-6595

\bibitem{teunissen2016}
Teunissen J and Ebert U 2016 {\em Plasma Sources Science and Technology\/} {\bf
  25} 044005 ISSN 0963-0252, 1361-6595

\bibitem{plewa2018a}
Plewa J~M, Eichwald O, Ducasse O, Dessante P, Jacobs C, Renon N and Yousfi M
  2018 {\em Journal of Physics D: Applied Physics\/} {\bf 51} 095206 ISSN
  0022-3727

\bibitem{roache1998}
Roache P~J 1998 {\em Verification and {{Validation}} in {{Computational
  Science}} and {{Engineering}}\/} 0th ed ({Albuquerque, N.M}: {Hermosa Pub})
  ISBN 978-0-913478-08-0

\bibitem{bagheri2018}
Bagheri B, Teunissen J, Ebert U, Becker M~M, Chen S, Ducasse O, Eichwald O,
  Loffhagen D, Luque A, Mihailova D, Plewa J~M, {van Dijk} J and Yousfi M 2018
  {\em Plasma Sources Science and Technology\/} {\bf 27} 095002 ISSN 1361-6595

\bibitem{pancheshnyi2005a}
Pancheshnyi S, Nudnova M and Starikovskii A 2005 {\em Physical Review E\/} {\bf
  71} ISSN 1539-3755, 1550-2376

\bibitem{ono2020}
Ono R and Komuro A 2020 {\em Journal of Physics D: Applied Physics\/} {\bf 53}
  035202 ISSN 0022-3727, 1361-6463

\bibitem{teunissen2017}
Teunissen J and Ebert U 2017 {\em Journal of Physics D: Applied Physics\/} {\bf
  50} 474001 ISSN 0022-3727, 1361-6463

\bibitem{bagheri2020a}
Bagheri B, Teunissen J and Ebert U 2020 {\em Plasma Sources Science and
  Technology\/} {\bf 29} 125021 ISSN 0963-0252, 1361-6595

\bibitem{francisco2021}
Francisco H, Bagheri B and Ebert U 2021 {\em Plasma Sources Science and
  Technology\/} ISSN 0963-0252

\bibitem{briels2008a}
Briels T~M~P, Kos J, Winands G~J~J, {van Veldhuizen} E~M and Ebert U 2008 {\em
  Journal of Physics D: Applied Physics\/} {\bf 41} 234004 ISSN 0022-3727,
  1361-6463

\bibitem{luque2008}
Luque A, Ratushnaya V and Ebert U 2008 {\em Journal of Physics D: Applied
  Physics\/} {\bf 41} 234005 ISSN 0022-3727, 1361-6463

\bibitem{komuro2012}
Komuro A, Ono R and Oda T 2012 {\em Journal of Physics D: Applied Physics\/}
  {\bf 45} 265201 ISSN 0022-3727, 1361-6463

\bibitem{Komuro_2013}
Komuro A, Ono R and Oda T 2013 {\em Plasma Sources Science and Technology\/}
  {\bf 22} 045002 ISSN 1361-6595
  \urlprefix\url{http://dx.doi.org/10.1088/0963-0252/22/4/045002}

\bibitem{komuro2018}
Komuro A, Matsuyuki S and Ando A 2018 {\em Journal of Physics D: Applied
  Physics\/} {\bf 51} 445204 ISSN 0022-3727, 1361-6463

\bibitem{eichwald2008}
Eichwald O, Ducasse O, Dubois D, Abahazem A, Merbahi N, Benhenni M and Yousfi M
  2008 {\em Journal of Physics D: Applied Physics\/} {\bf 41} 234002 ISSN
  0022-3727

\bibitem{nijdam2016}
Nijdam S, Teunissen J, Takahashi E and Ebert U 2016 {\em Plasma Sources Science
  and Technology\/} {\bf 25} 044001 ISSN 0963-0252, 1361-6595

\bibitem{marode2016}
Marode E, Dessante P and Tardiveau P 2016 {\em Plasma Sources Science and
  Technology\/} {\bf 25} 064004 ISSN 1361-6595

\bibitem{brisset2019}
Brisset A, Gazeli K, Magne L, Pasquiers S, Jeanney P, Marode E and Tardiveau P
  2019 {\em Plasma Sources Science and Technology\/} {\bf 28} 055016 ISSN
  1361-6595

\bibitem{zhu2021}
Zhu Y, Chen X, Wu Y, Hao J, Ma X, Lu P and Tardiveau P 2021 {\em Plasma Sources
  Science and Technology\/} ISSN 0963-0252

\bibitem{tholin2011}
Tholin F, Rusterholtz D~L, Lacoste D~A, Pai D~Z, Celestin S, Jarrige J, Stancu
  G~D, Bourdon A and Laux C~O 2011 {\em IEEE Transactions on Plasma Science\/}
  {\bf 39} 2254--2255 ISSN 1939-9375

\bibitem{pechereau2014a}
Pechereau F, Le~Delliou P, J{\'a}nsk{\'y} J, Tardiveau P, Pasquiers S and
  Bourdon A 2014 {\em IEEE Transactions on Plasma Science\/} {\bf 42}
  2346--2347 ISSN 1939-9375

\bibitem{yousfi2012}
Yousfi M, Eichwald O, Merbahi N and Jomaa N 2012 {\em Plasma Sources Science
  and Technology\/} {\bf 21} 045003 ISSN 0963-0252

\bibitem{hofmans2020}
Hofmans M, Viegas P, Rooij O~v, Klarenaar B, Guaitella O, Bourdon A and Sobota
  A 2020 {\em Plasma Sources Science and Technology\/} {\bf 29} 034003 ISSN
  0963-0252 publisher: IOP Publishing
  \urlprefix\url{https://doi.org/10.1088/1361-6595/ab6d49}

\bibitem{Viegas_2020}
Viegas P, Hofmans M, van Rooij O, Obrusník A, L~M~Klarenaar B, Bonaventura Z,
  Guaitella O, Sobota A and Bourdon A 2020 {\em Plasma Sources Science and
  Technology\/} {\bf 29} 095011 ISSN 1361-6595
  \urlprefix\url{http://dx.doi.org/10.1088/1361-6595/aba7ec}

\bibitem{li2012a}
Li C, Teunissen J, Nool M, Hundsdorfer W and Ebert U 2012 {\em Plasma Sources
  Sci. Technol.\/}  15

\bibitem{markosyan2015}
Markosyan A~H, Teunissen J, Dujko S and Ebert U 2015 {\em Plasma Sources
  Science and Technology\/} {\bf 24} 065002 ISSN 0963-0252, 1361-6595

\bibitem{dujko2013}
Dujko S, Markosyan A~H, White R~D and Ebert U 2013 {\em Journal of Physics D:
  Applied Physics\/} {\bf 46} 475202 ISSN 0022-3727, 1361-6463

\bibitem{nijdam2011}
Nijdam S, Wormeester G, {van Veldhuizen} E~M and Ebert U 2011 {\em Journal of
  Physics D: Applied Physics\/} {\bf 44} 455201 ISSN 0022-3727, 1361-6463

\bibitem{nijdam2014}
Nijdam S, Takahashi E, Markosyan A~H and Ebert U 2014 {\em Plasma Sources
  Science and Technology\/} {\bf 23} 025008 ISSN 0963-0252, 1361-6595

\bibitem{teunissen2018}
Teunissen J and Ebert U 2018 {\em Computer Physics Communications\/} {\bf 233}
  156--166 ISSN 00104655

\bibitem{zheleznyak1982}
Zheleznyak M, Mnatsakanyan A and Sizykh S 1982 {\em High Temperature\/} {\bf
  20} 357--362 ISSN 0018-151X

\bibitem{bourdon2007}
Bourdon A, Pasko V~P, Liu N~Y, C{\'e}lestin S, S{\'e}gur P and Marode E 2007
  {\em Plasma Sources Science and Technology\/} {\bf 16} 656--678 ISSN
  0963-0252, 1361-6595

\bibitem{bagheri2019}
Bagheri B and Teunissen J 2019 {\em Plasma Sources Science and Technology\/}
  {\bf 28} 045013 ISSN 1361-6595

\bibitem{phelps1985}
Phelps A~V and Pitchford L~C 1985 {\em Physical Review A\/} {\bf 31} 2932--2949
  ISSN 0556-2791

\bibitem{hagelaar2005}
Hagelaar G~J~M and Pitchford L~C 2005 {\em Plasma Sources Science and
  Technology\/} {\bf 14} 722--733 ISSN 0963-0252, 1361-6595

\bibitem{pancheshnyi2000a}
Pancheshnyi S~V, Sobakin S~V, Starikovskaya S~M and Starikovskii A~Y 2000 {\em
  Plasma Physics Reports\/} {\bf 26} 1054--1065 ISSN 1562-6938

\bibitem{BOLSIG+}
{{BOLSIG}}+ solver ver. 03/2016 www.lxcat.net

\bibitem{Phelps-database}
Phelps database ({{N2}},{{O2}}) www.lxcat.net, retrieved on January 19, 2021

\bibitem{xiaoran-psst-2020}
Li X, Sun A, Zhang G and Teunissen J 2020 {\em Plasma Sources Science and
  Technology\/} {\bf 29} 065004 ISSN 0963-0252

\bibitem{hansen1985}
Hansen E~W and Law P~L 1985 {\em Journal of the Optical Society of America A\/}
  {\bf 2} 510 ISSN 1084-7529, 1520-8532

\bibitem{schafer2011}
Schafer R 2011 {\em IEEE Signal Processing Magazine\/} {\bf 28} 111--117 ISSN
  1053-5888

\bibitem{naidis2009}
Naidis G~V 2009 {\em Physical Review E\/} {\bf 79} ISSN 1539-3755, 1550-2376

\bibitem{kossyi1992}
Kossyi I~A, Kostinsky A~Y, Matveyev A~A and Silakov V~P 1992 {\em Plasma
  Sources Science and Technology\/} {\bf 1} 207--220 ISSN 0963-0252, 1361-6595

\bibitem{pancheshnyi2013}
Pancheshnyi S 2013 {\em Journal of Physics D: Applied Physics\/} {\bf 46}
  155201 ISSN 0022-3727

\bibitem{aleksandrov1999}
Aleksandrov N~L and Bazelyan E~M 1999 {\em Plasma Sources Science and
  Technology\/} {\bf 8} 285--294 ISSN 0963-0252

\bibitem{stephens2018}
Stephens J 2018 {\em Journal of Physics D: Applied Physics\/} {\bf 51} 125203
  ISSN 0022-3727

\bibitem{tejero-del-caz2019}
{Tejero-del-Caz} A, Guerra V, Gon{\c c}alves D, da~Silva M~L, Marques L,
  Pinh{\~a}o N, Pintassilgo C~D and Alves L~L 2019 {\em Plasma Sources Science
  and Technology\/} {\bf 28} 043001 ISSN 0963-0252

\bibitem{biagi1999}
Biagi S~F 1999 {\em Nuclear Instruments and Methods in Physics Research Section
  A: Accelerators, Spectrometers, Detectors and Associated Equipment\/} {\bf
  421} 234--240 ISSN 0168-9002

\bibitem{rabie2016}
Rabie M and Franck C~M 2016 {\em Computer Physics Communications\/} {\bf 203}
  268--277 ISSN 0010-4655

\bibitem{Pancheshnyi_2012}
Pancheshnyi S, Biagi S, Bordage M, Hagelaar G, Morgan W, Phelps A and Pitchford
  L 2012 {\em Chemical Physics\/} {\bf 398} 148--153 ISSN 03010104

\bibitem{carbone2021}
Carbone E, Graef W, Hagelaar G, Boer D, Hopkins M~M, Stephens J~C, Yee B~T,
  Pancheshnyi S, van Dijk J and Pitchford L 2021 {\em Atoms\/} {\bf 9} 16
  number: 1 Publisher: Multidisciplinary Digital Publishing Institute
  \urlprefix\url{https://www.mdpi.com/2218-2004/9/1/16}

\bibitem{alves2014}
Alves L~L 2014 {\em Journal of Physics: Conference Series\/} {\bf 565} 012007
  ISSN 1742-6596 publisher: IOP Publishing
  \urlprefix\url{https://doi.org/10.1088/1742-6596/565/1/012007}

\bibitem{loureiro1986}
Loureiro J and Ferreira C~M 1986 {\em Journal of Physics D: Applied Physics\/}
  {\bf 19} 17--35 ISSN 0022-3727

\bibitem{IST-Lisbon-database}
Ist lisbon database ({{N2}},{{O2}}) www.lxcat.net, retrieved on January 19,
  2021

\bibitem{Morgan-database}
Morgan database ({{N2}},{{O2}}) www.lxcat.net, retrieved on January 19, 2021

\bibitem{TRINITI-database}
Triniti database ({{N2}},{{O2}}) www.lxcat.net, retrieved on January 19, 2021

\bibitem{Biagi-database}
Biagi database ({{N2}},{{O2}}) www.lxcat.net, retrieved on January 19, 2021

\bibitem{petrovic2009}
Petrovi{\'c} Z~L, Dujko S, Mari{\'c} D, Malovi{\'c} G, Nikitovi{\'c} {\v Z},
  {\v S}a{\v s}i{\'c} O, Jovanovi{\'c} J, Stojanovi{\'c} V and
  {Radmilovi{\'c}-Ra{\textbackslash}djenovi{\'c}} M 2009 {\em Journal of
  Physics D: Applied Physics\/} {\bf 42} 194002 ISSN 0022-3727

\bibitem{pancheshnyi2005}
Pancheshnyi S 2005 {\em Plasma Sources Science and Technology\/} {\bf 14}
  645--653 ISSN 0963-0252, 1361-6595

\end{thebibliography}

\end{document}